\documentclass[reprint,superscriptaddress]{revtex4-1}
\usepackage{subfigure}
\usepackage{graphicx}
\usepackage{natbib}
\usepackage{amsmath}
\usepackage{amssymb}
\usepackage{bbold}
\usepackage{xcolor}
\usepackage{colortbl}

\newcommand{\curlyE}{\mathcal{E}}
\newcommand{\curlyP}{\mathcal{P}}
\newcommand{\curlyEtilde}{\tilde{\mathcal{E}}}
 
 \newcommand{\dw}{\delta \omega}
 \newcommand{\Jth}{J_{\rm th}}
  \newcommand{\Jsb}{J_{\rm sb}}
  \newcommand{\dwsb}{\delta \omega_{\rm sb}}

\begin{document}

\title{Single-mode instability in standing-wave lasers: The quantum cascade laser as a self-pumped parametric oscillator}

\author{Tobias S. Mansuripur}
\affiliation{Department of Physics, Harvard University, Cambridge, MA 02138 USA}

\author{Camille Vernet}
\affiliation{John A. Paulson School of Engineering and Applied Sciences, Harvard University,  Cambridge, MA 02138 USA}
\affiliation{Ecole Polytechnique, 91123 Palaiseau, France}

\author{Paul Chevalier}
\affiliation{John A. Paulson School of Engineering and Applied Sciences, Harvard University,  Cambridge, MA 02138 USA}

\author{Guillaume Aoust}
\affiliation{John A. Paulson School of Engineering and Applied Sciences, Harvard University,  Cambridge, MA 02138 USA}
\affiliation{ONERA, The French Aerospace Lab, 91123 Palaiseau, France}


\author{Benedikt Schwarz}
\affiliation{John A. Paulson School of Engineering and Applied Sciences, Harvard University,  Cambridge, MA 02138 USA}
\affiliation{Institute of Solid State Electronics, TU Wien, 1040 Vienna, Austria}


\author{Feng Xie}
\affiliation{Thorlabs Quantum Electronics (TQE), Jessup, MD 20794 USA}

\author{Catherine Caneau}
\affiliation{Corning, Inc., Corning, NY 14831 USA}

\author{Kevin Lascola}
\affiliation{Thorlabs Quantum Electronics (TQE), Jessup, MD 20794 USA}

\author{Chung-en Zah}
\affiliation{Thorlabs Quantum Electronics (TQE), Jessup, MD 20794 USA}

\author{David P. Caffey}
\affiliation{Daylight Solutions, Inc., San Diego, CA 92128 USA}

\author{Timothy Day}
\affiliation{Daylight Solutions, Inc., San Diego, CA 92128 USA}

\author{Leo J. Missaggia}
\affiliation{Massachusetts Institute of Technology, Lincoln Laboratory, Lexington, MA 02420 USA}

\author{Michael K. Connors}
\affiliation{Massachusetts Institute of Technology, Lincoln Laboratory, Lexington, MA 02420 USA}

\author{Christine A. Wang}
\affiliation{Massachusetts Institute of Technology, Lincoln Laboratory, Lexington, MA 02420 USA}

\author{Alexey Belyanin}
\affiliation{Department of Physics and Astronomy, Texas A \& M University, College Station, TX 77843 USA}

\author{Federico Capasso}
\email{capasso@seas.harvard.edu} 
\affiliation{John A. Paulson School of Engineering and Applied Sciences, Harvard University,  Cambridge, MA 02138 USA}

\begin{abstract}
We report the observation of a clear single-mode instability threshold in continuous-wave Fabry-Perot quantum cascade lasers (QCLs). The instability is characterized by the appearance of sidebands separated by tens of free spectral ranges (FSR) from the first lasing mode, at a pump current not much higher than the lasing threshold. As the current is increased, higher-order sidebands appear that preserve the initial spacing, and the spectra are suggestive of harmonically phase-locked waveforms. We present a theory of the instability that applies to all homogeneously-broadened standing-wave lasers. The low instability threshold and the large sideband spacing can be explained by the combination of an unclamped, incoherent Lorentzian gain due to the population grating, and a coherent parametric gain caused by temporal population pulsations that changes the spectral gain line shape. The parametric term suppresses the gain of sidebands whose separation is much smaller than the reciprocal gain recovery time, while enhancing the gain of more distant sidebands. The large gain recovery frequency of the QCL compared to the FSR is essential to observe this parametric effect, which is responsible for the multiple-FSR sideband separation. We predict that by tuning the strength of the incoherent gain contribution, for example by engineering the modal overlap factors and the carrier diffusion, both amplitude-modulated (AM) or frequency-modulated emission can be achieved from QCLs. We provide initial evidence of an AM waveform emitted by a QCL with highly asymmetric facet reflectivities, thereby opening a promising route to ultrashort pulse generation in the mid-infrared. Together, the experiments and theory clarify a deep connection between parametric oscillation in optically pumped microresonators and the single-mode instability of lasers, tying together literature from the last 60 years.
\end{abstract}

\maketitle

\section{Introduction}
In the last decade, significant efforts have spurred the understanding of high-Q optically-pumped microresonators. A monochromatic external pump beam is coupled to a mode of the microresonator, and at sufficient pump power the third-order $\chi^{(3)}$ Kerr nonlinearity, responsible for the intensity-dependent refractive index, couples the pumped mode to fluctuations at other frequencies, which leads to interesting physics. Starting from an initial demonstration of third-order optical parametric oscillation (OPO) \cite{Kippenberg2004,Matsko2005}, in which the pump beam provides sufficient parametric gain to allow a few pairs of sidebands to oscillate, this technique has been extended to generate wide-spanning frequency combs \cite{DelHaye2011a,Herr2012}, and most recently temporal solitons \cite{Herr2013,Saha2013}. The many degrees of freedom one can manipulate in these systems, such as the group velocity dispersion (GVD), the free spectral range of the resonator, the detuning of the pump frequency relative to the cold cavity mode that it pumps, and the pump power, among others, have provided a rich nonlinear optical playground to observe diverse physical phenomena.

A laser, much like an OPO, is an optical resonator in which circulating monochromatic light reaches high intensity, the difference being that the light is internally generated rather than externally injected. Furthermore, the very gain medium that allows for lasing, simultaneously provides a third-order nonlinearity, the population pulsation (PP) nonlinearity \cite{Lamb1964}. The PP nonlinearity is an intrinsic property of any two-level system that interacts with near-resonant amplitude-modulated (AM) light: the radiative transition rate between the states, and therefore the population of each state, is temporally modulated by the AM light, resulting in so-called population pulsations that act back on the light field in a nonlinear way. The laser therefore contains the two ingredients, high-intensity light and a non-linearity, necessary for parametric oscillation. Indeed, in the late 1960s the importance of PPs in determining the above-threshold spectral evolution of a homogeneously broadened, traveling-wave laser was realized. At the laser threshold, one mode--which we call the primary mode--begins to lase and as the current is increased the population inversion remains clamped to its threshold value. It was first thought that this clamping should prevent any other mode from reaching the oscillation threshold. This reasoning, however, neglects the fact that when a photon of a different frequency is spontaneously emitted in the presence of the primary lasing field, a beat note--i.e., an intensity modulation at the difference frequency of the two fields--is created. The beat note creates a PP that provides a parametric contribution to the gain of the spontaneously emitted photon. At a sufficiently high pumping level known as the instability threshold, this parametric gain can--despite the fact that the population inversion is clamped to its threshold value--allow two sidebands to overcome the loss. The separation of these sidebands from the primary mode is related to the Rabi frequency induced by the primary mode. This effect is responsible for both the Haken-Risken-Schmid-Weidlich (HRSW) instability \cite{Haken1966, Risken1966} and the Risken-Nummedal-Graham-Haken (RNGH) instability \cite{Risken1968a, Graham1968}. Many years later, insightful work properly identified the fundamental role of PPs in the single-mode laser instabilities \cite{Hendow1982a, Hendow1982b, Hillman1982, Lugiato1983, Hillman1984, Lugiato1985, Hillman1985} and also chaos \cite{Gioggia1983}. (We note that PPs are important not only for inverted media. Historically, their effects were first appreciated in microwave spectroscopy pump-probe experiments by Autler and Townes \cite{Autler1955} in 1955, and soon came to be known as the ac Stark effect. Through the late 1960s and 1970s, significant work on sideband amplification \cite{Senitzky1963}, resonance fluorescence \cite{Senitzky1968, Newstein1968}, and the Mollow scattering triplet \cite{Mollow1969a, Stroud1971, Schuda1974, Wu1975, Carmichael1976} culminated in the ``dressed" description of atoms in strong fields \cite{Cohen-Tannoudji1977}. In the 1980s, the PP nonlinearity was cast in the language of nonlinear optics and applications such as four-wave-mixing \cite{Boyd1981}, phase conjugation \cite{Harter1980}, and optical bistability \cite{Hillman1982} were explored.)

\begin{figure}
\includegraphics[scale=1.05]{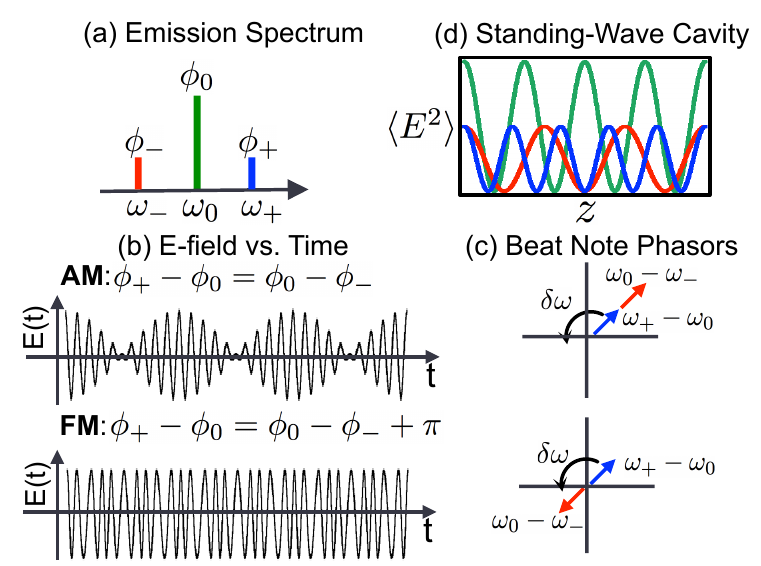}
\caption{\label{fig:AMvsFM}(a) The emission spectrum at the instability threshold comprises a primary mode and two weak sidebands.  (b) The temporal behavior of the field $E(t)$ depends on the relative phases of the three modes $\phi_-$, $\phi_0$, and $\phi_+$, and shown are the AM and FM configurations. (c) The AM and FM fields can be understood in terms of the constructive and destructive addition of two beat note phasors, where each phasor represents a contribution to the intensity modulation at the difference frequency $\dw$ resulting from the superposition of each sideband with the primary mode. (d) In a standing-wave cavity, the intensity of each mode varies with position, and the spatial modes corresponding to different frequencies do not perfectly overlap.}
\end{figure}

Both the HRSW and RNGH single-mode instabilities apply to homogeneously-broadened traveling-wave lasers, and predict the appearance of sidebands on the primary lasing mode, as shown in Fig.\ \ref{fig:AMvsFM}(a). We remark that in general, the temporal behavior of an electric field that contains three equally spaced frequencies can be more amplitude-modulated or frequency-modulated (FM), depending on the spectral phase, as shown in Fig.\ \ref{fig:AMvsFM}(b). One can think of the intensity modulation (in other words, the beat note) of the AM and FM fields as resulting from the sum of two phasors rotating at frequency $\dw$, each of which is created by the beat between a sideband and the primary mode. As shown in Fig.\ \ref{fig:AMvsFM}(c), the two phasors either constructively interfere to create a large intensity modulation (AM) or destructively interfere to eliminate the intensity modulation (FM). In both the HRSW and RNGH instabilities, the three-wave field is by necessity AM; a constant-intensity FM field would not create the PP and the resulting parametric gain that is required by the sidebands to reach the lasing threshold. In the HRSW case, which applies to low quality-factor cavities for which the photon lifetime is shorter than the atomic decay time, the sideband separation is smaller than the mode spacing, or free spectral range (FSR), of the cavity. All three lasing frequencies fall within a single cold cavity resonance, which is made possible by a region of anomalous dispersion created by the PP \cite{Hendow1982a}. In the RNGH instability, which applies to higher quality-factor cavities, the sidebands must coincide with cold cavity modes in order to satisfy the roundtrip phase condition, resulting in a separation that is an integer multiple of the FSR. An important corollary of this requirement is that to observe the effect of the PPs, the FSR must be smaller than the gain recovery frequency (i.e., inverse of the gain recovery time $T_1$). Why? The gain recovery time determines the shortest time scale at which the population inversion can respond to an intensity modulation; therefore, the amplitude of the PP is only significant for sidebands detuned by an amount close to or smaller than $1/T_1$. If the FSR is greater than $1/T_1$, then all cavity modes are too far from the primary mode to generate a PP with an amplitude large enough to make the sidebands unstable. Ideally, the FSR should be significantly smaller than $1/T_1$ so that the FP modes densely populate the parametric gain lobe, increasing the probability of satisfying the instability condition. Provided this condition is met, the RNGH instability predicts that a traveling-wave laser with rapid dephasing must be pumped nine times above threshold before the instability appears. Experimental observations of a rhodamine dye ring laser \cite{Hillman1984} showed signatures of an RNGH-like instability, with two key differences: the instability threshold was only fractionally higher, not nine times higher, than the lasing threshold, and the sideband creation was accompanied by the disappearance of the primary mode. Efforts to explain the discrepancies between theory and experiment are well-summarized in \cite{Lugiato1985,Lugiato1987}, but to our knowledge the discrepancy was never fully resolved.

In this work, we will investigate the single-mode instability in a standing-wave laser, shown schematically in Fig.\ \ref{fig:AMvsFM}(d). The distinguishing feature of the standing-wave laser is that the primary mode induces a population grating (PG) (as long as carrier diffusion is limited), an effect known as spatial hole burning (SHB). The gain of other cavity modes is no longer clamped above threshold, but continues to increase with the pumping. Therefore, the instability threshold can be reached {\em without} the need for PP parametric gain. We call this an incoherent instability, and it occurs in media whose gain recovery time is too slow for PPs to occur ($\mathrm{FSR}>\mathrm{1/T_1}$), such as diode lasers. In gain media with a fast recovery time ($\mathrm{FSR}<\mathrm{1/T_1}$), both the incoherent gain and the parametric PP contribution to the gain must be considered. We will show that the PP parametrically suppresses the gain of nearby sidebands, because low-frequency sidebands cause the population inversion to oscillate perfectly out of phase with the intensity modulation. On the other hand, the PP enhances the gain of larger-detuning sidebands, as occurs in the RNGH instability. Depending on the relative contributions of the incoherent and coherent gain, we show that the laser will either emit an FM or an AM waveform at the instability threshold, to either minimize or maximize the amplitude of the PPs. If the incoherent gain is large, nearby sidebands are favored and will yield FM emission to minimize the amount of parametric suppression. If the incoherent gain is small, larger-detuning sidebands are favored and will yield AM emission to maximize the amount of parametric enhancement. The possibility of both FM and AM emission from a standing-wave laser is a novelty not shared by the traveling-wave laser, which, as mentioned before, can only produce an AM waveform.
 
The quantum cascade laser (QCL) is precisely the kind of laser for which both the PG and PPs are important. An electron injected into the upper state has only a short picosecond lifetime during which to diffuse before it scatters to the ground state--not enough time to traverse the half-wavelength mid-infrared ($\lambda \sim 3$\,-12\,$\mu$m) standing-wave from node to antinode. Therefore, the PG is not washed out. Also, the FSR (typically 8 to 16\,GHz) is much less than the gain recovery frequency ($1/T_1 \approx 1$\,THz), so the population inversion has no difficulty following the beat notes in field intensity created when multiple modes lase simultaneously, yielding PPs. We report the discovery that continuous-wave (cw) Fabry-Perot (FP) QCLs reach a well-defined instability threshold, characterized by the appearance of sidebands whose separation from the primary mode can be several multiples of the cavity FSR. This mode skipping is a clear signature of the parametric PP interaction between the primary mode and the sidebands, which strongly suppresses sidebands at separations much smaller than the large gain recovery frequency of the QCL. The behavior is observed in QCLs that emit at wavelength 3.8\,$\mu$m, 4.6\,$\mu$m, and 9.8\,$\mu$m, indicating that it is a universal feature of mid-infrared QCLs, independent of the specific bandstructure of the active region. The strength of the PG can be tuned by coating the facets to adjust their reflectivities. By comparing the measurements with the theory, we argue that QCLs with uncoated facets emit an FM waveform. A QCL with one high-reflectivity facet and a sufficiently low reflectivity of the other facet should in principle emit an AM waveform, and we provide preliminary evidence that this is indeed the case, demonstrating a QCL whose sidebands are separated from the primary mode by 46 FSR. While the PG and PP have been known to be important in QCLs, in previous work their effects were treated separately \cite{Gordon2008}. Instead, we emphasize that  one should think of the PG--a spatial modulation of the inversion--and the PP--a temporal modulation of the inversion--as working {\em in tandem} to create a phase-locked multimode state at low pump power.

As the current is increased past the instability threshold, higher-order sidebands that preserve the initial spacing appear. This suggests that the FP-QCL can emit a harmonically phase-locked waveform without the need for any external modulation or additional nonlinear elements. Why have such spectra not been observed before, except in a few cases \cite{Tober2014,Lu2015}? We have found the harmonic states to be extremely sensitive to optical feedback. Simply placing a collimating lens between the QCL and the spectrometer--even a poorly aligned, tilted lens with a focal length of a few cm--makes it difficult to observe the harmonic state, and instead yields the more familiar QCL spectrum in which all adjacent FP modes lase. It is also important to slowly increase the current, which allows for a smooth transition from the single-mode to the harmonic regime. We argue that the harmonic state is an intrinsic regime of all QCLs. The fact that it has only been observed 15 years after the invention of the cw QCL is a testament to the destabilizing influence of optical feedback \cite{Soriano2013}. 

In the last few years, comb generation in a QCL on adjacent FP modes has been demonstrated \cite{Hugi2012, Villares2014, Villares2016}, and the importance of parametric mode coupling is known \cite{Khurgin2014, Villares2015}. (These devices all had multi-stage inhomogeneously broadened active regions, which distinguishes them from the devices in our work.) Because these combs have so far always comprised adjacent cavity modes (with the exception of \cite{Li2015}), consideration has only been given to the case where the fundamental frequency of the PPs equals the FSR. This low PP frequency strongly favors the emission of an FM waveform. The remarkable degree of freedom to skip modes, never before considered, means that the temporal periodicity of the PPs is no longer pinned to the cavity roundtrip time (typically 60 to 120\,ps), but is shortened by a factor equal to the number of modes skipped, which reaches 46 in one of our QCLs. This reduction of the period down to the order of the gain recovery time is the crucial feature that allows for the possibility of AM emission.

Finally, we emphasize the deep connection between the single-mode laser instability and mode proliferation in optically pumped microresonators. Both are cases of parametric oscillation that are initiated by a nonlinearity, either PP or Kerr, transferring energy from a pump beam to two sidebands. For a passive microresonator the pump beam must be injected, while in the laser the pump beam is internally generated. This analogy, which we only begin to uncover here, can help guide future work toward understanding the rich emission spectra of QCLs. More broadly, both QCLs and microresonators exhibit the widespread phenomenon of modulation instability \cite{Zakharov2009}. We hope that the advancement of the QCL can parallel the rapid progress seen in microresonators in the last decade, leading to a compact source of mid-infrared frequency combs for spectroscopy of trace gases and short pulse generation \cite{Schliesser2012}.

In Sec.\ \ref{sec:Experiment} we present the experimental results, which helps to motivate the theory presented in Sec.\ \ref{sec:Theory}. In Sec.\ \ref{sec:Discussion} we compare the theory with the measurements, and finally conclude in Sec.\ \ref{sec:Conclusion}.

\section{Experiment} \label{sec:Experiment}
\begin{figure}
\includegraphics[scale=0.97]{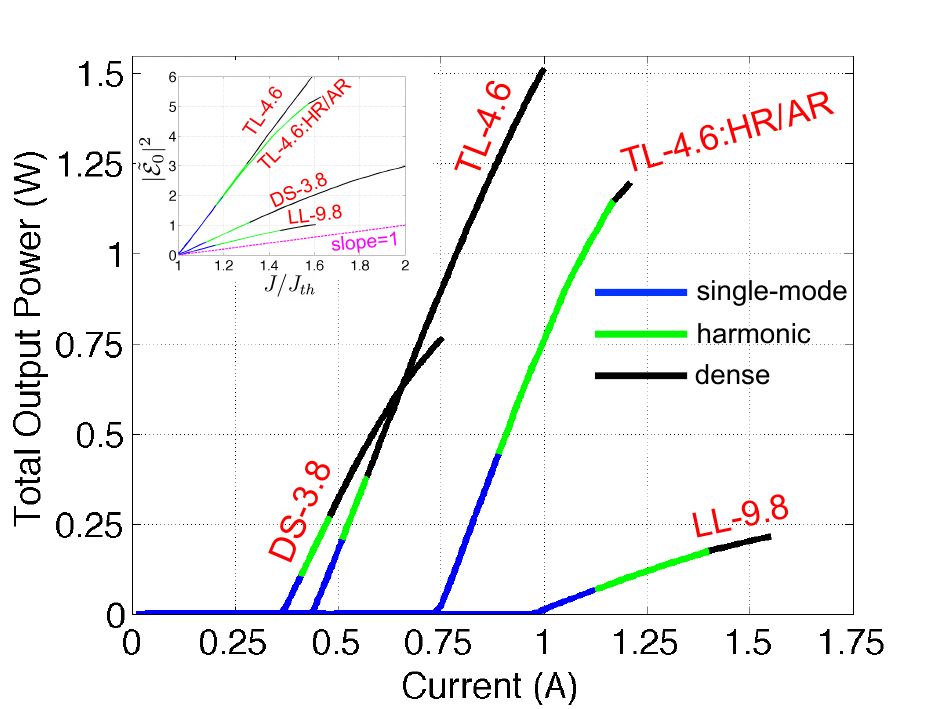}
\caption{\label{fig:LI}Total power output of each QCL (from both facets) vs. current, color-coded to indicate the range over which the laser operates in a single-mode (the region following threshold), harmonic state (the middle region), or dense state (the highest-power region). Inset: the intracavity power normalized to the saturation intensity (calculated from the measured output power and the best estimates for $\kappa$, $T_1$, $T_2$, and the facet reflectivities) is plotted vs. $J/J_{\rm th}$.}
\end{figure}

All four devices used in this study are cw, buried heterostructure, FP-QCLs. Our device naming convention identifies the provider of the device (LL: MIT Lincoln Laboratory, TL: Thorlabs, DS: Daylight Solutions) followed by the emission wavelength in microns. The active region of device LL-9.8 is a double phonon resonance design using lattice-matched Ga$_{0.47}$In$_{0.53}$As/Al$_{0.48}$In$_{0.52}$As, grown by metalorganic chemical vapor deposition, with the well-known layer structure of \cite{Hofstetter2001} (with a nominal doping of $n=2.5 \times 10^{18}$\,cm$^{-3}$), for which extensive bandstructure calculations have been done \cite{Faist2002}. The device length is 3\,mm and width is 8\,$\mu$m. Devices TL-4.6, TL-4.6:HR/AR, and DS-3.8 were grown using strained Ga$_x$In$_{1-x}$As/Al$_y$In$_{1-y}$As and are described in \cite{Xie2011}, although the layer sequence is not given. The length is 6\,mm and width is 5\,$\mu$m for these three devices. Both facets are left uncoated for LL-9.8, TL-4.6, and DS-3.8. The only coated device is TL-4.6:HR/AR, which has a high-reflectivity (HR) coating on the back facet ($R\approx 1$) and an antireflection (AR) coating on the front facet ($R \approx 0.01$), but is otherwise nominally identical to TL-4.6. Far-field measurements indicate that all devices exhibit single lateral-mode emission over the full range of applied current. It is worth mentioning that the short-wave QCLs, DS-3.8 and TL-4.6, have positive GVD and the long-wave device LL-9.8 has negative GVD. We expect this because their wavelengths lie on opposite sides of the zero-GVD point of InP, but we have also confirmed this in Appendix \ref{app:sec:GVD} using the subthreshold measurement method of \cite{Hofstetter1999}. Some relevant parameters for each device are given in Table \ref{tab:deviceparameters}: the effective refractive index $n_{\rm eff}$ is determined from the FP-mode spacing of the measured spectra; the dipole moment $d$ and the upper state lifetime $T_{\rm up}$ are calculated from the bandstructure; the dephasing time $T_2$ is determined from a Lorentzian fit to either an electroluminescence or far-subthreshold measurement of each device. The output power of each laser was measured with a calibrated thermopile (Ophir 3A-QUAD) placed close to the facet; the total output power is plotted in Fig.\ \ref{fig:LI}, which is obtained from the front facet only for TL-4.6:HR/AR and by doubling the single-facet power of the uncoated lasers. 

\begin{center}
\begin{table}
\resizebox{\columnwidth}{!}{
\begin{tabular}{|l|c|c|c|c|c|c|c|c|}
\hline
Device & $n_{\rm eff}$ & d [nm$\cdot e]$ & $T_{\rm up} $ [ps] & $\gamma_D$ & $T_2$ [fs] & $\dw_{\rm FSR}$ [GHz] & $\dwsb$ [GHz] & $\Jsb/\Jth$ \\ \hline
LL-9.8 & 3.43 & 3 & 0.54 & 0.93 & 81 & 92 & 642  & 1.14 \\ \hline
TL-4.6 & 3.23 & 1.63 & 1.7 & 0.49 & 74 & 48 & 1259  & 1.17 \\ \hline
TL-4.6:HR/AR & 3.25 & 1.63 & 1.7 & 0.49 & 74 & 48 & 2216  & 1.22 \\ \hline
DS-3.8 & 3.25 & 1.5 & 1.74 & 0.40 & 43 & 49 & 977  & 1.12 \\
\hline
\end{tabular}}
\caption{Summary of relevant parameters of the devices used in this study. $n_{\rm eff}$, $T_2$, $\dwsb$, and $\Jsb$ were measured quantities. $T_{\rm up}$ and $d$ were calculated from the bandstructure for TL-4.6 and DS-3.8, and taken from \cite{Faist2002} for LL-9.8, and $\gamma_D$ was calculated assuming $D=77$\,cm$^2$/s \cite{Faist2013}. \label{tab:deviceparameters} }
\end{table}
\end{center}

Our goal was to precisely examine the spectral evolution of the QCL with increasing current, from the single-mode to the multimode regime. Specifically, we wanted to answer the question: at what pumping level does a second mode start to lase, and what is the relationship between the second frequency and the first? To answer this question, we would begin each measurement with the laser driven at a current beneath the laser threshold. The current was then slowly increased in steps of 1\,mA (at a rate of roughly 2 mA per second), and the spectrum was monitored using a Fourier transform infrared (FTIR) spectrometer (Bruker Vertex 80v), with either an InSb detector (for DS-3.8 and TL-4.6) or HgCdTe detector (for LL-9.8), both cryogenically cooled. The current was supplied by a low-noise driver (Wavelength Electronics QCL1500 or QCL2000), and the temperature of the copper block beneath the QCL was stabilized to 15$^\circ$C. The slow rate of increase of the current was necessary to precisely identify the instability threshold, and also to prevent rapid temperature variations. To completely eliminate the possibility of optical feedback due to reflections from optical elements outside the laser cavity, the QCL was placed about 40\,cm from the entrance window to the FTIR and its output was not collimated with a lens, but simply allowed to diverge. The high power of the devices and the sensitivity of the detectors was sufficient to measure spectra despite the small fraction of collected optical power. 

\begin{figure}
\includegraphics[scale=0.97]{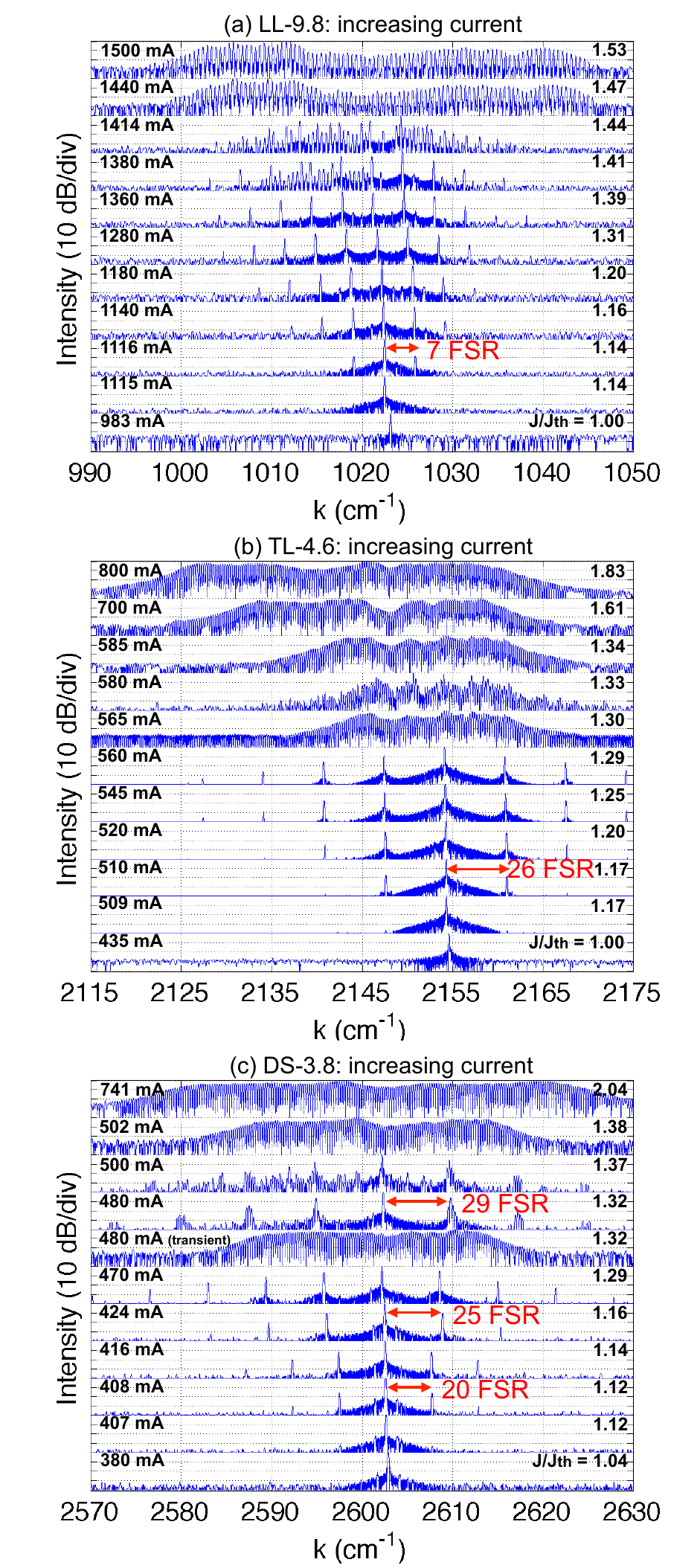}
\caption{\label{fig:spectra_uncoated}Spectra of the three uncoated QCLs (a) LL-9.8, (b) TL-4.6 and (c) DS-3.8 as the current is incremented, starting from below threshold.}
\end{figure}

Spectra measured in this manner are shown in Fig.\ \ref{fig:spectra_uncoated} for the three uncoated devices. Each spectrum is normalized to its own maximum and plotted on a logarithmic scale covering 40\,dB of intensity variation. All three lasers undergo a very similar spectral evolution. Above threshold, the laser remains single-mode for a substantial range of current until a clear instability threshold is reached, at which a 1\,mA increase in current results in the appearance of new lasing modes. The new frequencies appear as symmetric sidebands on the primary lasing frequency, with a separation that is many integer multiples of the FSR. The sideband spacing $\dw_{\rm sb}$ and pumping $J_{\rm sb}$ at the sideband instability threshold are given in Table \ref{tab:deviceparameters} for each device. Taking LL-9.8 as a first example, at $J_{\rm sb}/J_{\rm th} = 1.14$ a pair of equal-amplitude sidebands separated by 7 FSR from the primary mode suddenly rise out of the noise floor to an intensity 20 dB weaker than the primary mode. As the current increases further, higher-order sidebands appear that preserve the initial spacing, eventually yielding a spectrum at $J/J_{\rm th}=1.39$ of 11 modes, each separated by 7 FSR from its nearest neighbors. We refer to a spectrum of modes separated by multiple FSR as a harmonic state. Above $J/J_{\rm th}=1.39$, interleaving modes incommensurate with the harmonic spacing begin to appear. At $J/J_{\rm th}=1.47$, there is another sudden transition at which all adjacent FP modes are populated; we refer to this as a ``dense" state, and it persists for all higher currents. For device TL-4.6, sidebands with a separation of 26 FSR from the primary mode appear at $J_{\rm sb}/J_{\rm th} = 1.17$, and the transition to the dense state occurs at $J/J_{\rm th}=1.30$. (In a second device nominally identical to TL-4.6, sidebands separated by 13 FSR appeared at $J_{\rm sb}/J_{\rm th} = 1.19$ and the dense state appeared at $J/J_{\rm th} = 1.31$.) For device DS-3.8, the sideband separation is 20 FSR at $J_{\rm sb}/J_{\rm th}= 1.12$. As the current is increased, the sideband spacing displays a sudden jump from 20 FSR to 25 FSR. At $J/J_{\rm th}=1.32$ the laser jumps to a dense state for somewhere between a few seconds and a minute before returning to a ``noisy" harmonic state: one with prominent harmonic peaks but many incommensurate modes populated as well. At $J/J_{\rm th}=1.38$ the dense state appears again, and this time persists for all higher currents.

When the spectral evolution measurement is repeated many times for one device, starting from below threshold and incrementing the current, we find that the instability threshold $J_{\rm sb}$ and sideband spacing $\dw_{\rm sb}$ are always the same. As the current is increased past $J_{\rm sb}$, there can be slight variations from one experiment to another. For example, the jump from 20 to 25 FSR in DS-3.8 does not always occur at the exact same current, but predictably within a range of about 20\,mA. The same is true of the transition to the dense state.

\begin{figure}
\includegraphics[scale=0.97]{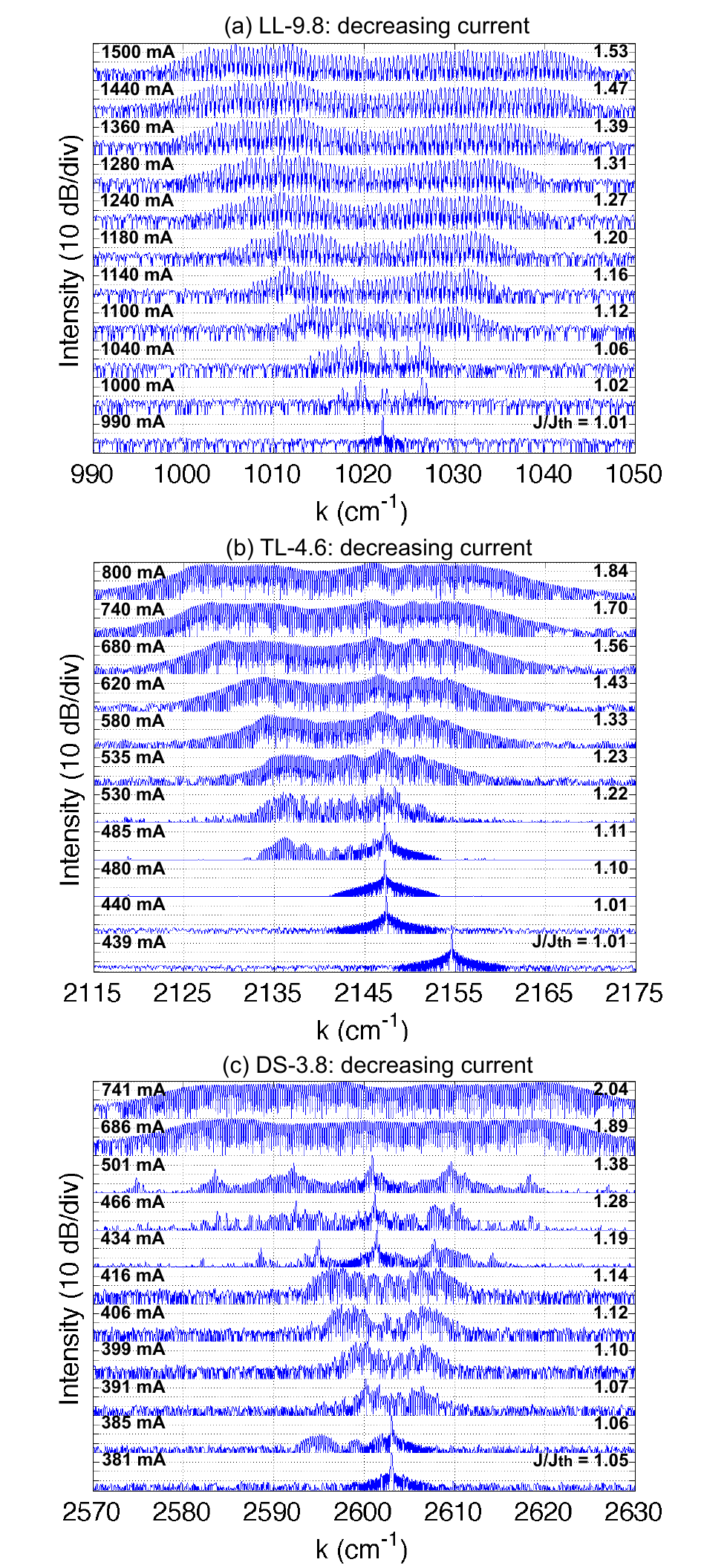}
\caption{\label{fig:spectra_uncoated_rampdown}Spectra of the three uncoated QCLs (a) LL-9.8, (b) TL-4.6 and (c) DS-3.8 as the current is decremented, starting from the current reached at the end of the upward current ramp shown in Fig.\,\ref{fig:spectra_uncoated}.}
\end{figure}

After the laser enters the dense state, we decrease the current slowly and observe a remarkable hysteresis in the spectral evolution, shown in Fig.\,\ref{fig:spectra_uncoated_rampdown}. In LL-9.8, the dense state persists all the way until $J/J_{\rm th}=1.01$, when the single-mode finally reappears. In TL-4.6, the dense state gives way to a single-mode at 2148 cm$^{-1}$ at $J/J_{\rm th}=1.10$, and then at  $J/J_{\rm th}=1.01$ jumps to a single-mode at 2155 cm$^{-1}$, which is the same mode observed at threshold when the current is ramped up in Fig.\,\ref{fig:spectra_uncoated}(b). In DS-3.8, a noisy harmonic state appears at $J/J_{\rm th}=1.38$, then laser returns to the dense state at $J/J_{\rm th}=1.14$, and the single-mode state reappears at $J/J_{\rm th}=1.05$. We emphasize the general observation for all three devices that the clean harmonic state cannot be recovered once the current has been increased far into the dense state regime. (If the ramp down is begun from a current not too much larger than the one at which the laser enters the dense state, harmonic states can reappear.) Additionally, there is no spectral hysteresis in the immediate vicinity of the instability threshold; for example, if DS-3.8 is toggled between 407 and 408 mA, then the spectrum simply toggles between the two spectra shown in Fig.\,\ref{fig:spectra_uncoated}(c), and the same is true for LL-9.8 and TL-4.6.

\begin{figure}
\includegraphics[scale=0.43]{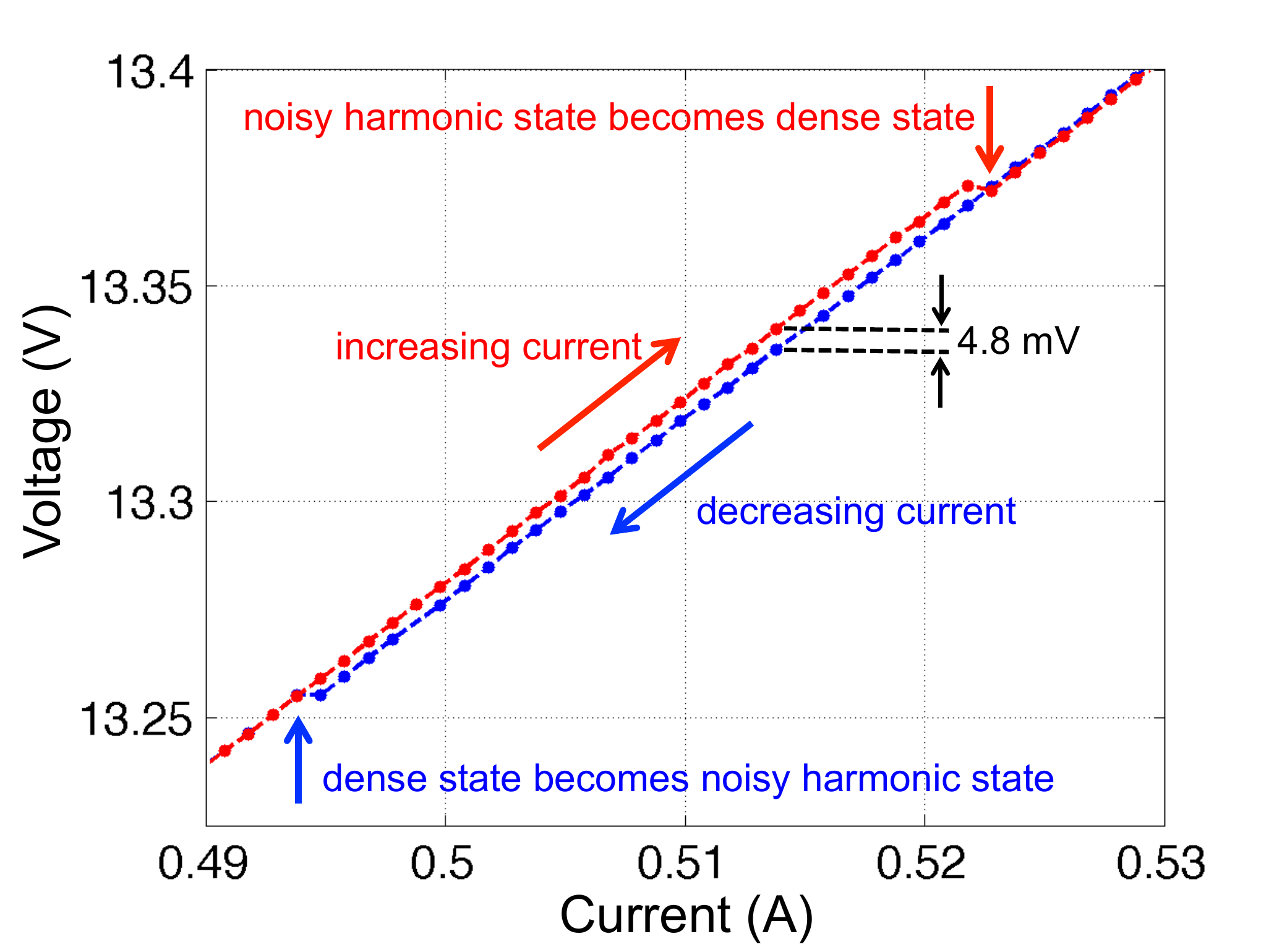}
\caption{\label{fig:IVhysteresis}The IV curve of DS-3.8 exhibits a hysteresis as the current is increased (red, upper curve) and decreased (blue, lower curve). The hysteresis is correlated with the transition from the noisy harmonic state to the dense state on the ramp up, and from the dense state to the noisy harmonic state on the ramp down.}
\end{figure}

One might expect sudden changes in the emission spectrum to be accompanied by changes in the output power and voltage. Since we can more sensitively measure the voltage than the output power, we plot in Fig.\,\ref{fig:IVhysteresis} a portion of the IV curve of DS-3.8. When starting below threshold and increasing the current, the voltage of the laser decreases (negative differential resistance) when the noisy harmonic state transitions to the dense state at 523 mA. (Note that in the spectra shown in Fig.\ 3(c), this transition occurred at 502 mA.) The IV curve also exhibits a hysteresis correlated with the spectral hysteresis: as the current is decreased after reaching the dense state, the laser remains in the dense state. For the same current, the voltage is 4.8 mV lower in the dense state than in the noisy harmonic state. (Accordingly, the output power is slightly higher in the dense state.) At 494 mA, the dense state transitions to the noisy harmonic state, and the two voltage curves overlap again.

\begin{figure*}
\includegraphics[scale=0.97]{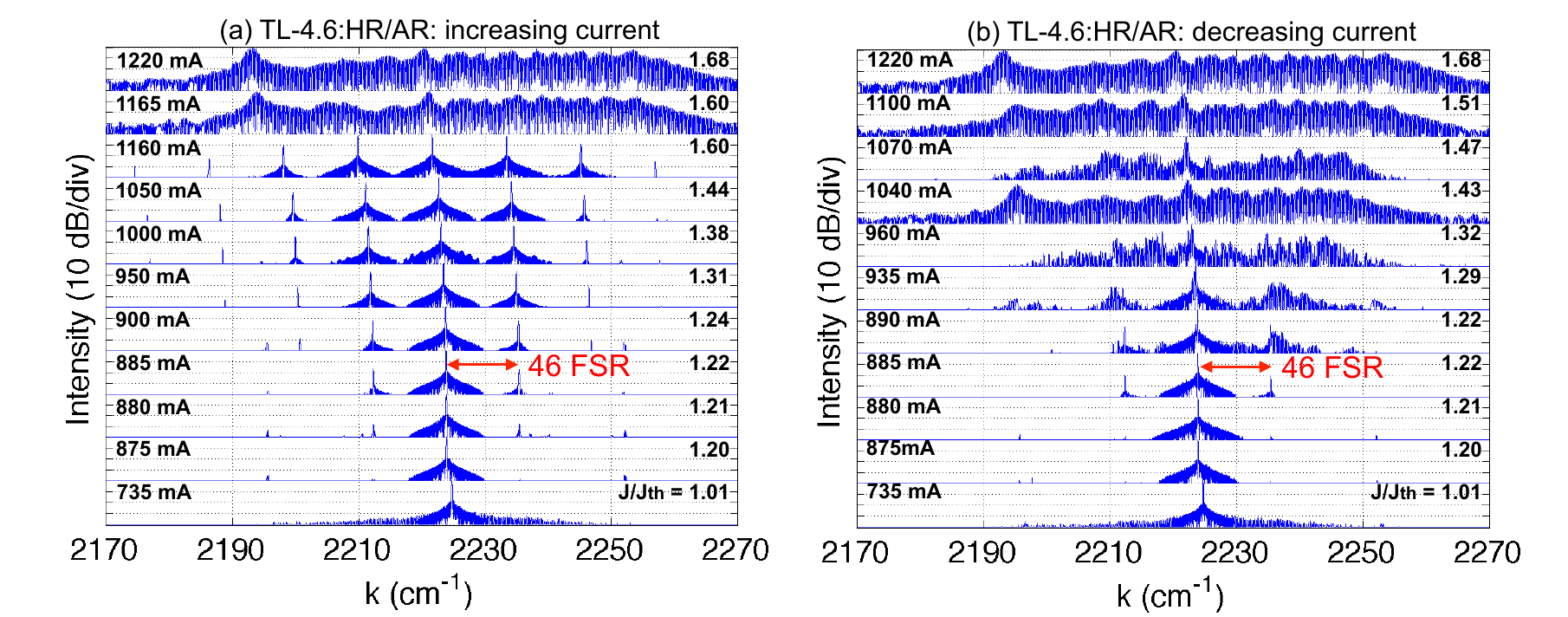}
\caption{\label{fig:TL-4.6:HR/AR}Spectra of TL-4.6:HR/AR as the current is (a) incremented, starting from below threshold, and (b) decremented, starting from the current reached at the end of the upward current ramp in (a).}
\end{figure*}

Lastly, we present in Fig.\ \ref{fig:TL-4.6:HR/AR}(a) the spectral evolution of TL-4.6:HR/AR  as the current is incremented. The behavior of this device is different from the uncoated devices in two significant ways: 1) the sidebands appear with a separation of 46 FSR, much larger than any spacing seen previously, and 2) the harmonic regime persists over a much larger range of output power--from $J/J_{\rm th} = 1.22$ to 1.60,  than it does in the uncoated devices, as seen by the color-coding in Fig.\ \ref{fig:LI}. (A second device, nominally identical to TL-4.6:HR/AR, developed sidebands with a spacing of 48 FSR at $J_{\rm sb}/J_{\rm th} = 1.18$ and also remained in the harmonic state over a large current range.) As the current is decreased, shown in Fig.\ \ref{fig:TL-4.6:HR/AR}(b), the clean harmonic state with one pair of sidebands reappears at 885 mA, which is quite close to the instability threshold of 880 mA found when the current is ramped upwards. The single-mode state reappears at $J/J_{\rm th} = 1.20$. 

To better understand the coherence properties of a multimode state, it is typical to look at the width of the inter-mode radio-frequency beat note generated on a photodetector. We present some measurements of the beat note in the dense state in Appendix \ref{app:sec:beatnote}. We find a range of interesting phenomena at various currents \cite{Villares2016}, including narrow beat notes, multiple closely-spaced beat notes, and broad beat notes. However, the focus of this paper is not on the coherence properties of the dense state \cite{Faist2016}, but rather to understand the transition from the single-mode to the harmonic state. Unfortunately, we cannot perform beat note measurements of the harmonic state because the smallest observed beat frequency is greater than 100\,GHz, larger than the electrical bandwidth of any mid-infrared photodetector. The observed spectra of the harmonic state strongly suggest, however, that the modes are phase-locked with one another through a nonlinear interaction, because it would be difficult to explain the mode-skipping if each lasing frequency acted as an independent oscillator. In the following theory section, we assume that the primary mode is coherent with the two sidebands and find that the consequences are consistent with our observations. The theory predicts that the uncoated devices emit FM waveforms, and suggests that the HR/AR device emits an AM waveform. In future work, second-order autocorrelation measurements are needed to experimentally verify this prediction.

\section{Theory} \label{sec:Theory}
The instability threshold is characterized by the appearance of symmetric sidebands on the primary lasing mode. Our goal is to theoretically explain the frequency separation of the sidebands and the pump power at which they first appear. We begin with the general framework: the Maxwell-Bloch equations for a two-level system and the spatial mode expansion of a laser cavity. Then, we address the single-mode solution of the laser to determine how the primary mode and the population grating (PG) evolve with increasing pumping. Next, we understand how a two-level system responds to a weak field at a frequency different from that of the primary mode--the population pulsation (PP). Finally, we will combine these two ingredients, the PG and the PP, to explain the instability threshold. We find that the PG provides an unclamped Lorentzian contribution to the gain of the sidebands, which is responsible for the low instability threshold. The PP reshapes the gain, suppressing nearby sidebands and enhancing more distant ones, and is responsible for the observed multiple-FSR sideband separation. Interestingly, we find that depending on the relative contributions of the PG and the PP to the gain, the laser can emit either an FM or AM waveform at the instability threshold.

\subsection{General Framework}
We model the lasing transition as a two-level system, or a quantum dipole, subject to the electric field
\begin{equation}
E(t) = \curlyE(t) e^{i\omega_0 t} + c.c.
\end{equation}
The response is characterized by the population inversion $w$ (positive when inverted) and the off-diagonal element of the density matrix $\sigma$, which in turn obey the Bloch equations (in the rotating wave approximation) \cite{Allen1987},
\begin{align}
\dot{\sigma} &= \left( i \Delta - \frac{1}{T_2} \right) \sigma + \frac{i \kappa}{2} w \curlyE \label{eq:Bloch_sigma} \\
\dot{w} &= i \kappa (\curlyE^* \sigma - \curlyE \sigma^*) - \frac{w - w_{eq}}{T_1}. \label{eq:Bloch_w}
\end{align}
where $\Delta = \omega_{ba} - \omega_0$ is the detuning between the field and the resonant frequency $\omega_{ba}$ of the two-level system, $T_1$ is the gain recovery time, $T_2$ is the dephasing time, $\kappa \equiv 2d/\hbar$ is the coupling constant where $d$ is the dipole matrix element (assumed to be real) and $\hbar$ is Planck's constant, and $w_{eq}$ is the ``equilibrium" population inversion that the system would reach in the absence of photons, determined by the pumping. (Note that we have defined $T_1$ to be the gain recovery time, which in QCLs is distinct from the upper state lifetime $T_{\rm up}$ due to the nature of electron transport in the active region. See Appendix \ref{app:sec:TupvsT1} for a discussion of this point.) We write the macroscopic polarization $P$ (dipole moment per volume) as
\begin{equation}
P(t)= \curlyP e^{i\omega_0 t} + c.c.,
\end{equation}
where $\curlyP = N d \sigma$, and $N$ is the volume density of dipoles.

A characteristic of the two-level medium that will appear often is the ``Beer rate"
\begin{equation}
\bar{\alpha} = \frac{N d^2 T_2 \omega_{ba} c \sqrt{\mu/\epsilon}}{\hbar},
\end{equation}
with dimensions of frequency, which is related to the more familiar Beer coefficient $\alpha$ (with dimensions of inverse length) that appears in Beer's law of absorption by $\bar{\alpha} = \alpha c$. The Beer rate gives the amount of loss when the material is in its ground state ($w=-1$), and also the maximum amount of gain when the material is fully inverted ($w=1$). We adopt the convention of \cite{Siegman1986} and assume our dipoles to be embedded in a host medium of permittivity $\epsilon$ and permeability $\mu$. The speed of light $c=1/\sqrt{\epsilon \mu}$ also denotes the value in the background host medium. 

In the laser cavity, the field envelopes vary in space and time. One approach is to numerically solve the full spatiotemporal Maxwell-Bloch equations \cite{Wojcik2013,Wang2015}. To obtain analytical results, we follow a common approximation and decouple the spatial and temporal dependencies \cite{Chembo2010}, writing the field as
\begin{equation}
E(z,t) = \sum_{m=-,0,+} \Upsilon_m(z) \curlyE_m(t) e^{i \omega_m t} + c.c.,
\end{equation}
where $\omega_0$ is the primary mode frequency and the sideband frequencies are $\omega_\pm = \omega_0 \pm \dw$. These three frequencies are cold-cavity resonant frequencies, and are equidistant from one another because we have assumed zero GVD. We will henceforth assume that the primary mode $\omega_0$ lases at the resonant frequency of the two-level system, so $\Delta = 0$. This is a reasonable approximation if the FSR is much smaller than the gain bandwidth. These two assumptions, $\mathrm{GVD}=0$ and $\Delta=0$, simplify later mathematical formulas considerably and allow for easier understanding of the essential physics. The full theory without these assumptions is included in Appendix \ref{app:sec:PPs}. The spatial modes $\Upsilon_m(z)$ are determined by the cavity geometry, and do not vary in time. We assume a linear FP cavity with two end mirrors of unity reflectivity as shown in Fig.\,\ref{fig:AMvsFM}(d), so that the spatial modes are given by
\begin{equation}
\Upsilon_m(z) = \sqrt{2} \cos(k_m z)
\end{equation}
where $k_m$ is an integer multiple of $\pi/L$ and $L$ is the length of the cavity. (The simpler case of the ring cavity is included in Appendix \ref{app:sec:instabilitythreshold}.) The mirror loss  $\ln (1/\sqrt{R_1 R_2})/L$ is included in the total optical losses of the cavity $\bar{\ell}$. The assumption of perfect reflectivity simplifies the problem in two important ways: the spatial functions $\Upsilon_m(z)$ are orthogonal, and they do not change shape as the pumping increases. This assumption turns out to be quite good even for semiconductor lasers with facet reflectivities around 0.25. The approximation breaks down for our HR/AR coated QCL, so the theory will only be directly applied to the uncoated lasers, but the implications of the theory for the HR/AR device will be qualitatively discussed.

\subsection{Population Grating}
\label{sec:PG}
The threshold inversion is given by the ratio of the optical loss rate to the Beer rate, $w_{th} = \bar{\ell}/\bar{\alpha}$. We define the pumping parameter $p \equiv w_{eq}/w_{th}$. When $p=1$, the primary mode begins to lase at the frequency $\omega_0 = \omega_{ba}$. As the pumping $p$ is increased, the field and inversion can be solved for by the method of \cite{Gordon2008} as shown in Appendix \ref{app:sec:singlemodetheory}, the main results of which are stated here. We account for the population grating, but not the coherence grating which has been incorporated in recent work \cite{Vukovic2016a}. The primary mode $\curlyEtilde_0$ grows according to
\begin{equation}
|\curlyEtilde_0|^2 = \frac{p-1}{1+\gamma_D/2}, \label{eq:LI}
\end{equation}
where we have defined the dimensionless primary mode amplitude $\curlyEtilde_0$ by normalizing by the saturation amplitude, $\curlyEtilde_0 \equiv \kappa \sqrt{T_1 T_2} \curlyE_0$. The diffusion parameter $\gamma_D$ is given by $\gamma_D = (1+4k_0^2 D T_{\rm up})^{-1}$, where $D$ is the lateral diffusivity of the excited-state electrons and $T_{\rm up}$ is the upper-state lifetime. The parameter $\gamma_D$ ranges from 0 (for infinite mobility) to 1 (for zero mobility). The population inversion in the presence of the primary mode, $w_0(z)$, varies with $p$ as
\begin{equation}
w_0(z) = w_{th} \left[ 1 + \frac{\gamma_D}{2} \frac{p-1}{1 + \gamma_D/2} - \gamma_D \frac{p-1}{1+\gamma_D/2} \cos(2k_0 z) \right] \label{eq:w0(z)_standingwave}.
\end{equation}
Equations \ref{eq:LI} and \ref{eq:w0(z)_standingwave} are valid to first order in the primary mode intensity $|\curlyEtilde_0|^2$, or equivalently, $p-1 \ll 1$. Note that for zero diffusion ($\gamma_D=1$), the slope efficiency of the laser is two thirds that of the infinite diffusion ($\gamma_D=0$) case \cite{Danielmeyer1971}. This is because for infinite diffusion, the inversion is uniformly pinned to $w_{th}$ above threshold. For finite diffusion, as the pumping increases the population grating grows in amplitude. At the same time, the average value of the inversion increases, indicating that the inversion is not being converted into photons as efficiently as it could be if the electrons could diffuse from the field nodes to the antinodes. In this derivation, we have assumed that the pump parameter $p$ is constant along the length of the laser; in an efficient electrically-pumped laser more current will flow to the field antinodes, which will reduce the amplitude of the population grating.

In principle, one can extract $\gamma_D$ from measurements of the slope of $|\curlyEtilde_0|^2$ vs. $p$, which should be between zero and one. The inset of Fig.\ \ref{fig:LI} shows $|\curlyEtilde_0|^2$ vs. $J/J_{\rm th}$. All curves have a slope greater than one, which suggests that $J/J_{\rm th}$ is an underestimate of $p$. (See Appendix \ref{app:sec:intracavitypower} for how $|\curlyEtilde_0|^2$ is determined from the measurements, and how a non-zero transparency current causes $J/J_{\rm th}$ to underestimate $p$.) Therefore, more characterization is needed to extract $\gamma_D$ from the measurements.

\subsection{Population Pulsation}
\label{sec:PP}
To understand the population pulsation, we can ignore the spatial dependence of the intracavity field and consider only a single two-level system subject to an applied field
\begin{equation}
E(t) = \sum_{m=-,0,+} \curlyE_m(t) e^{i \omega_m t} + c.c.
\end{equation}
Since we will later be interested in calculating the stability of the sidebands, the amplitudes $\curlyE_\pm$ should be thought of as infinitesimal perturbations; as such, our entire treatment retains only terms to first order in the sideband amplitudes $\curlyE_\pm$. Full details of the calculation are in Appendix \ref{app:sec:PPs}. We write the total polarization as
\begin{equation}
P(t) = \sum_{m=-,0,+} \curlyP_m(t) e^{i \omega_m t} + c.c.
\end{equation}
The polarization at the sidebands can be calculated using Eqs. \ref{eq:Bloch_sigma} and \ref{eq:Bloch_w} \cite{Boyd2003}, which gives
\begin{align}
\curlyP_+ &= \frac{i \epsilon}{\omega_{ba}} \bar{\alpha} w_0 \left[ \frac{\curlyE_+}{[1 + i  \dw T_2]} + \Lambda \curlyEtilde_0 \curlyEtilde_0^*  \curlyE_+ + \Lambda \curlyEtilde_0 \curlyEtilde_0 \curlyE_-^*\right] \label{eq:P+} \\
\curlyP_- &= \frac{i \epsilon}{\omega_{ba}} \bar{\alpha} w_0 \left[ \frac{\curlyE_-}{[1 - i  \dw T_2]} + \Lambda^* \curlyEtilde_0 \curlyEtilde_0^*  \curlyE_- + \Lambda^* \curlyEtilde_0 \curlyEtilde_0 \curlyE_+^*\right] , \label{eq:P-}
\end{align}
where
\begin{equation}
\Lambda = \frac{ - (1+i \dw T_2/2) }{\left[ (1 + i \dw T_1) (1 + i \dw T_2)^2  + (1+ i\dw T_2) |\curlyEtilde_0|^2 \right] } \label{eq:Lambda++_Delta=0}
\end{equation}
and $w_0$ is the saturated population inversion $w_0 = w_{eq}/(1 + |\curlyEtilde_0|^2)$. 

The polarization at each sideband is neatly divided into three contributions.  Taking $\curlyP_+$ as an example, the first term in Eq.\ \ref{eq:P+} is the Lorentzian contribution that the sideband generates due to the linear susceptibility of the dipole. The second and third terms are nonlinear contributions due to the PP at frequency $\dw$: a self-mixing term of the sideband with the primary mode, and a cross-mixing term of the other sideband with the primary mode. The frequency-dependent portion of the nonlinear susceptibility is $\Lambda$, which is a dimensionless function of the sideband detuning, the time constants of the two-level system, and the primary mode intensity $|\curlyEtilde_0|^2$. From the field and the induced polarization, we can calculate the total power density generated, $\langle - E \dot{P}\rangle$. The quantity that most interests us is the gain $\bar{g}$ (with dimension of frequency) seen by each sideband, defined as the power generated at the sideband's frequency, divided by the energy density of the exciting sideband field. 

To develop a feel for the parametrically generated polarization and the resulting gain, we consider two instructive cases. In both cases we take the sidebands to have equal magnitudes, $|\curlyE_+|=|\curlyE_-|$, but choose the phases of the sidebands to give rise to an AM waveform in one case and a constant-intensity FM waveform in the other case, as shown in Fig.\ \ref{fig:AMvsFM}(b). The gain of each sideband is found to be
\begin{equation}
\bar{g} = \bar{\alpha} w_0 \left[ \frac{1}{1+ (\dw T_2)^2} + {\rm Real}(\Lambda) |\curlyEtilde_0|^2 \times \left\{ \label{eq:sidebandgain_travelingwave}
\begin{array}{cc}
2 & {\rm ;\ AM} \\
0 & {\rm ;\ FM}
\end{array}
\right.
\right].
\end{equation}
The first term is the Lorentzian contribution to the gain, and the second term is the parametric gain due to the PP. The factors of two and zero come from the constructive or destructive addition, respectively, of the self-mixing and cross-mixing terms to the nonlinear polarization. Equivalently, one can say that the constant-intensity FM field does not create a PP, and accordingly experiences no parametric gain. The parametric gain of the AM field is proportional to ${\rm Real}(\Lambda)$ and to the primary mode intensity $|\curlyEtilde_0|^2$. (We note that one can quickly derive the original RNGH instability for a traveling-wave laser from Eq.\ \ref{eq:sidebandgain_travelingwave}, which is shown in Appendix \ref{app:sec:instabilitythreshold}.) By expanding $\Lambda$ in Eq.\ \ref{eq:Lambda++_Delta=0} in powers of $|\curlyEtilde_0|^2$, it becomes clear that the PP interaction can be expressed in the perturbative expansion of traditional nonlinear optics as a third, fifth, seventh, etc. order nonlinearity. We will later calculate the instability threshold in the limit of small primary mode intensity, and are therefore interested in the lowest-order nonlinearity. We obtain $\chi^{(3)}$, the dimensionless frequency-dependent portion of the third-order PP nonlinear susceptibility, by evaluating $\Lambda$ at $|\curlyEtilde_0|^2=0$,
\begin{equation}
\chi^{(3)} = \frac{ - (1+i \dw T_2/2) }{(1 + i \dw T_1) (1 + i \dw T_2)^2  } \label{eq:Xi3}.
\end{equation}


\begin{figure}
\includegraphics[scale=0.9]{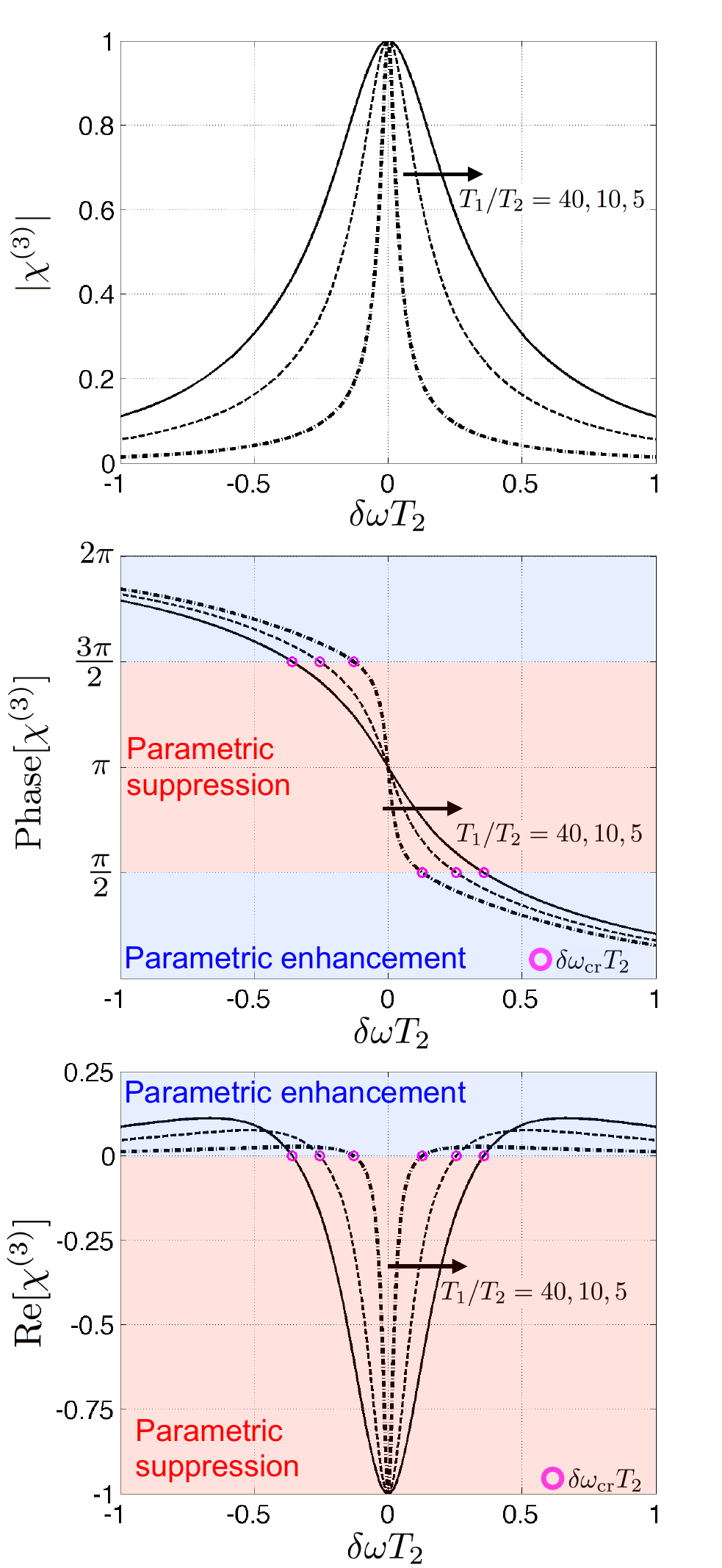}
\caption{\label{fig:chi3} The magnitude, phase, and real part of $\chi^{(3)}$ are plotted vs. $\dw T_2$ for three different ratios $T_1/T_2=40,10,5$. The parametric gain seen by the sidebands is determined by ${\rm Real}[\chi^{(3)}]$. Low-frequency PPs lead to a parametric suppression of the gain. For the gain to be parametrically enhanced, $\dw$ must be large enough that the inversion can no longer follow the intensity in anti-phase; in other words, the phase of $\chi^{(3)}$ must be between $-\pi/2$ and $\pi/2$. }
\end{figure}


To better elucidate the nature of the PPs, the magnitude, phase, and real part of $\chi^{(3)}$ are plotted in Fig.\ \ref{fig:chi3} as a function of the sideband detuning, for three different values of $T_1/T_2$. At low frequencies $\dw$, the population inversion has no difficulty following the modulation of the field, which has two consequences: the amplitude of the PPs is large, and the PP is $\pi$ out of phase with the intensity modulation of the exciting field. This can be understood simply in terms of rate equations: when the field is stronger, the stimulated emission rate is larger, and the population inversion is therefore smaller. This scenario--higher inversion when the intensity is lower and lower inversion when the intensity is higher--is less efficient at extracting power from the two-level system relative to the case of monochromatic or FM excitation; mathematically, this is described by a parametric gain (determined by the real part of $\chi^{(3)}$) that is negative. We refer to this effect as parametric suppression: a low-frequency PP reduces the gain of each sideband. As $\dw$ increases, the inversion can no longer as easily follow the intensity modulation, so the amplitude of the PPs decreases and the phase of $\chi^{(3)}$ decreases from $\pi$. For large enough $\dw$, the phase of $\chi^{(3)}$ decreases below $\pi/2$, at which point ${\rm Real}[\chi^{(3)}]$ becomes positive. We refer to this effect as parametric enhancement: a high-frequency PP increases the gain seen by each sideband. The crossing frequency $\dw_{\rm cr}$ which separates the low-frequency suppression regime and high-frequency enhancement regime is given by
\begin{equation}
\dw_{\rm cr} T_2 \approx \sqrt{\frac{2/3}{T_1/T_2}}, \label{eq:dw_cr}
\end{equation} 
where we have made the approximation $T_1/T_2 \gg 1$, valid for QCLs. The regions of parametric suppression and enhancement are highlighted in the plots of the phase and real part of $\chi^{(3)}$ in Fig.\ \ref{fig:chi3}. Finally, at very large $\dw$ the parametric gain approaches zero (from above), because the beat note becomes too short for the inversion to follow and the amplitude of the PP approaches zero.


It is worth pointing out that in the weak-field limit $|\curlyEtilde_0|^2 \ll 1$ that we are interested in, $\dw_{\rm cr}$ has no relation to the Rabi frequency $\Omega_R$ induced by the primary mode, 
\begin{equation}
\Omega_R T_2 = \frac{|\curlyEtilde_0|}{\sqrt{T_1/T_2}}. \label{eq:Omega_R}
\end{equation}
The Rabi frequency of course varies with the primary mode amplitude, while $\dw_{\rm cr}$ is independent of $\curlyEtilde_0$ in the weak-field limit. By comparing the factors $\sqrt{2/3}$ and $|\curlyEtilde_0|$ in the numerators of Eqs.\ \ref{eq:dw_cr} and \ref{eq:Omega_R}, it's clear that in the limit $|\curlyEtilde_0|^2 \ll 1$, $\dw_{\rm cr}$ will always be greater than the Rabi frequency. Thus, the reason for the parametric enhancement when $\dw > \dw_{\rm cr}$ should simply be ascribed to the fact that at high PP frequency, the phase lag between the population inversion and the field intensity becomes appropriate for gain rather than absorption.



\subsection{Instability Threshold}
\label{sec:instabilitythreshold}
Now we can ask the question: what happens to the single-mode laser solution when it is perturbed by a weak sideband field? The source of the perturbation could be spontaneous emission, or even spontaneous parametric downconversion of two primary mode photons into two sideband photons \cite{Kleinman1968}. Our goal is to calculate the gain of the sideband modes averaged over the length of the cavity. The instability threshold is reached when the sideband gain equals the loss.  Although our instability analysis will not tell us about the steady-state reached by the sidebands, one reasonable possibility is that the sidebands begin to lase, as seen in the experimental spectra.


To determine the sideband gain, we start with the polarization in Eqs.\ \ref{eq:P+}-\ref{eq:P-} and account for the position-dependence by replacing $\curlyE_m(t)$ with $\curlyE_m(t) \Upsilon_m(z)$, and $w_0$ with $w_0(z)$ from Eq.\ \ref{eq:w0(z)_standingwave}. In keeping with our approximation to order $|\curlyEtilde_0|^2$, we replace $\Lambda$ with $\chi^{(3)}$. The position-dependent polarization is then inserted as the source term in Maxwell's wave equation. From here, the calculation follows the same steps as the instability analysis done for Kerr microresonators \cite{Chembo2010}, and is detailed in Appendix \ref{app:sec:instabilitythreshold}. After making the slowly varying envelope approximation and projecting the equation onto each of the orthonormal spatial modes, one finds a first order differential equation for each sideband amplitude. Unlike the earlier example where we hand-picked the phases of the sidebands to study the effect of an AM and FM field, here the AM and FM sideband configurations emerge organically as the two ``natural modes" of the system of two sideband equations. The natural modes \cite{Hillman1982} are the configurations of the three-wave field for which the relative phases of the fields are preserved as time evolves; in other words, an AM field remains AM, and an FM field remains FM. (In the general case of nonzero $\Delta$ and GVD, the natural modes can be a superposition of AM and FM.) The gain of the AM and FM natural modes is given by
\begin{align}
\frac{\bar{g}} {\bar{\ell}} &=  \frac{1 + \frac{\gamma_D}{2} \frac{p-1}{1+\gamma_D/2}}{1 + (\dw T_2)^2} \nonumber \\
&+ {\rm Real}[\chi^{(3)}]  \frac{p-1}{1+\gamma_D/2} \cdot \left\{
\begin{array}{cc}
\Gamma_{\rm self} + \Gamma_{\rm cross} = \frac{3}{2} & {\rm ;\ AM} \\
\Gamma_{\rm self} - \Gamma_{\rm cross} = \frac{1}{2} & {\rm ;\ FM}
\end{array}
\right. \label{eq:standingwave_sidebandgain}
\end{align}
where the $\Gamma$s are longitudinal spatial overlap factors
\begin{align}
\Gamma_{\rm self} &= \frac{1}{L} \int_0^L dz\ |\Upsilon_0(z)|^2 |\Upsilon_{\pm}(z)|^2 = 1 \\
\Gamma_{\rm cross}&= \frac{1}{L} \int_0^L dz\ \Upsilon_0(z)^2 \Upsilon_-^*(z) \Upsilon_+^*(z) = 1/2.
\end{align}
By comparing the standing-wave sideband gain in Eq.\ \ref{eq:standingwave_sidebandgain} to the sideband gain of a single two-level system in Eq.\ \ref{eq:sidebandgain_travelingwave}, we see that the cavity introduces two modifications. First, the Lorentzian gain contribution increases with $p$; this unclamped gain is a direct result of the PG that develops in the presence of non-zero $\gamma_D$. Secondly, the partial overlap of the sideband spatial modes $\Upsilon_+$ and $\Upsilon_-$ results in partial (rather than complete) interference of the self-mixing and cross-mixing contributions to the gain. To understand this, note from Fig.\ \ref{fig:AMvsFM}(a) that although the {\em emitted} waveform has equal-amplitude sidebands, {\em within} the cavity the plus and minus sidebands have unequal amplitudes at most positions $z$, as shown by the red and blue modes in Fig.\ \ref{fig:AMvsFM}(d). Therefore, the self and cross-mixing contributions to the sideband polarization at each position $z$ cannot completely interfere, and the factors of $3/2$ (AM) and $1/2$ (FM) emerge after averaging over the full cavity length, as opposed to the factors of 2 and 0 in Eq.\ \ref{eq:sidebandgain_travelingwave}. Thus, even when the laser emits an FM waveform, there is still a parametric contribution to the gain due to the incomplete destructive interference of the PP within the cavity.



\begin{figure}
\includegraphics[scale=0.9]{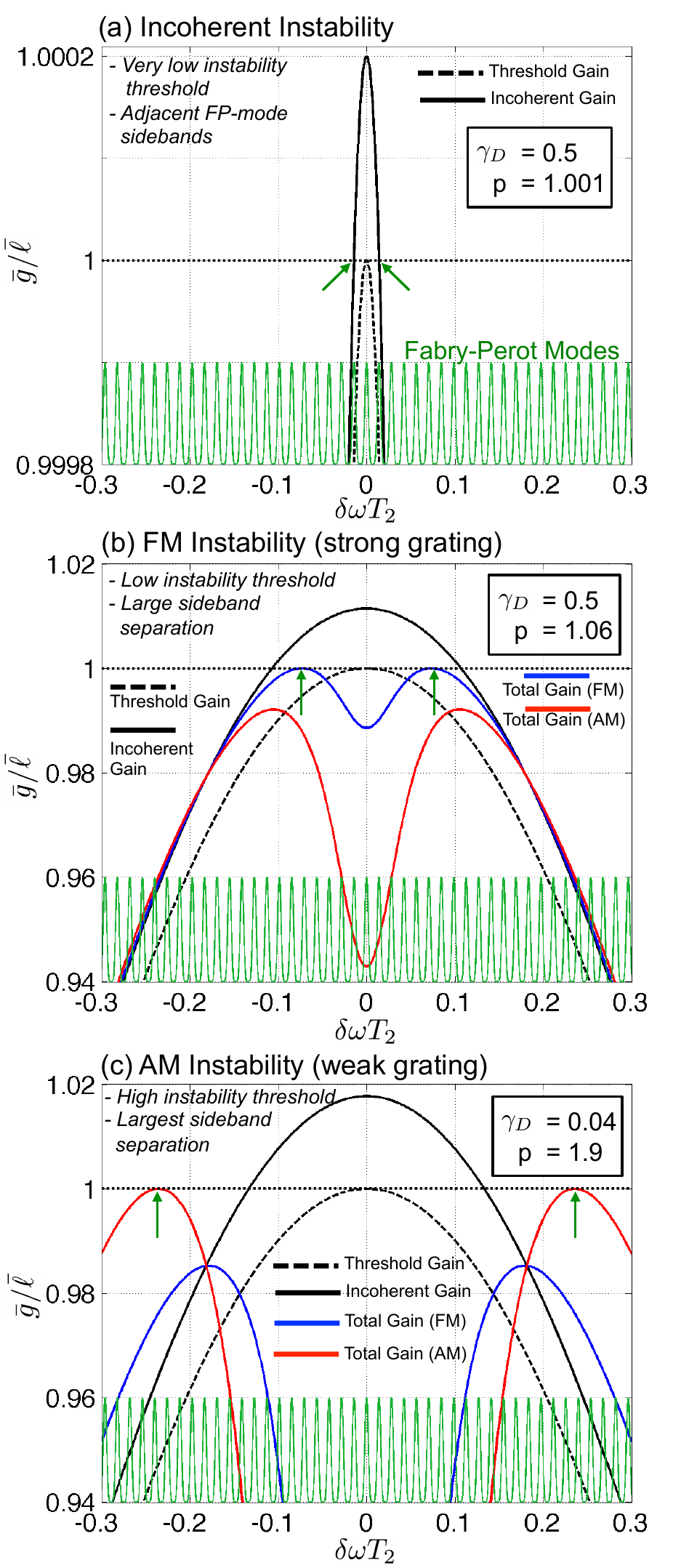}
\caption{\label{fig:typesofinstabilities} Overview of the three different types of instabilities. (a) The incoherent instability relies only on the unclamped gain due to the PG, and occurs when parametric effects can be neglected. (b) The FM instability occurs when the gain due to a strong PG, despite the parametric suppression of the gain of small-detuned sidebands, allows the less-suppressed FM sidebands to reach threshold.  (c) The AM instability occurs when the gain due to a weak PG, together with the parametric enhancement of the gain of large-detuned sidebands, allows the more strongly enhanced AM sidebands to reach threshold. In (b) and (c), the value $T_1/T_2=20$ was used. The FP modes (green, bottom of each panel) are not associated with the ordinate, and simply provide a sense of the mode spacing.}
\end{figure}

The instability occurs when $p$ reaches a value such that the sideband gain $\bar{g}$ in Eq.\ \ref{eq:standingwave_sidebandgain} equals the loss $\bar{\ell}$ for one particular sideband detuning $\dw$. (As discussed previously, we assume the FSR is small so that a FP mode always exists very close to the unstable value of $\dw$.) As $p$ increases, the incoherent Lorentzian gain increases, but the parametric gain either increases or becomes more negative depending on the sign of ${\rm Real}[\chi^{(3)}]$ (which depends on $\dw$).  The three parameters $T_1$, $T_2$, and $\gamma_D$ affect the relative importance of the incoherent and coherent gain terms, and depending on the values of these parameters, one of three different classes of instability can occur: the incoherent instability, FM instability, and AM instabilty. In Fig.\ \ref{fig:typesofinstabilities}, each type of instability is illustrated by plotting the sideband gain at the instability threshold, which we now explain.

\subsubsection{Incoherent Instability}
The parametric gain can often be ignored. If $T_1$ is large enough,  the interesting features of ${\rm Real} [\chi^{(3)}]$ all occur for sideband detunings less than 1 FSR, and so the parametric gain will be nearly zero for all values of $\dw$ greater than 1 FSR. This is the case for diode lasers, where PPs are significant up to a few GHz ($T_1 \approx 1$\,ns), while the FSR is around 100\,GHz. Thus, only the incoherent gain term in Eq.\ \ref{eq:standingwave_sidebandgain} matters (although it is not a Lorentzian for bandgap lasers). As $p$ increases beyond 1, the sideband gain increases but remains Lorentzian, so the sidebands that reach the instability threshold first will always be the FP modes immediately adjacent to the primary lasing mode \cite{Holonyak1980}. In diode lasers, $\gamma_D$ is small ($\sim 10^{-4}$), so $p$ needs to be large before the second mode can appear.

The value of $\gamma_D=0.5$ in Fig.\ \ref{fig:typesofinstabilities}(a) is typical of short-wave QCLs. We see that if coherent effects were negligible in QCLs, we would expect the sidebands to appear at $p=1.001$, barely above threshold. The much higher instability threshold measured in the experiments, together with the observation that the sidebands do not appear at the nearest-neighbor FP modes of the primary mode, indicates that coherent effects play an essential role in the QCL instability.

\subsubsection{FM Instability}
When the Lorentzian gain increases quickly with $p$ due to a strong PG, sidebands that fall within the parametric suppression regime, $\dw < \dw_{\rm cr}$, can reach the instability threshold. This is counterintuitive: why should a sideband lase when the parametric interaction provides negative gain? The answer is that the Lorentzian gain favors sidebands with as small a separation as possible, and if the Lorentzian gain is large enough it can pull sidebands above threshold in spite of the negative contribution from the parametric gain. In this scenario, FM sidebands have a lower instability threshold than AM sidebands because the parametric contribution to the gain is {\em less negative}, since $1/2 < 3/2$ in Eq.\ \ref{eq:standingwave_sidebandgain}. Such a case is illustrated in Fig.\ \ref{fig:typesofinstabilities}(b) for $\gamma_D=0.5$ and $T_1/T_2=20$. At p=1.06, FM sidebands reach the instability threshold, while AM sidebands are too strongly suppressed to reach the instability. A key feature of the instability is that the unstable sideband will be several FSR away from the primary mode (provided that the FSR is small), while still satisfying  $\dw < \dw_{\rm cr}$.

\subsubsection{AM Instability}
When the Lorentzian gain increases little with $p$ due to a weak PG, only sidebands that fall within the parametric enhancement regime, $\dw > \dw_{\rm cr}$, will be able to reach the instability. In this case, AM will have a lower instability threshold than FM because AM receives a larger parametric enhancement (since $3/2 > 1/2$ in Eq.\ \ref{eq:standingwave_sidebandgain}). Such a case is illustrated in Fig.\ \ref{fig:typesofinstabilities}(c) for $\gamma_D=0.04$ and $T_1/T_2=20$. At $p=1.9$, AM sidebands reach the instability threshold while the FM sidebands are not sufficiently enhanced to reach the instability. Strictly speaking, p=1.9 falls outside the region of validity of our perturbative treatment ($p-1 \ll 1$), so the specific values in this plot are not exactly accurate, but the qualitative features are correct. The unstable sidebands satisfy $\dw > \dw_{\rm cr}$, and so their separation will be even greater than for the FM instability. The original RNGH instability is precisely this AM instability, in a traveling-wave laser.  For traveling waves, the Lorentzian gain is clamped at threshold regardless of the diffusion parameter, so the instability can only be reached by the parametric enhancement of AM sidebands.

To access both the FM and AM instability regimes experimentally, we need to tune the strength of the PG. The electron diffusivity can be reduced by lowering the temperature, and indeed temperature has a strong effect on the emission spectra of QCLs \cite{Gordon2008}, although the effect is not yet well-understood. In this work, we choose to manipulate the PG by adjusting the facet reflectivities. Increasing the disparity of the reflectivities of the two mirrors reduces the contrast of the standing-wave, because the wave traveling from the higher to the lower-reflectivity facet becomes larger than the counter-propagating wave \cite{Mansuripur2015}. For a sufficiently large disparity, the incoherent gain contribution is small enough that the laser can only undergo the AM instability. In a practical sense, engineering the facet coatings allows one to transform a standing-wave cavity into more of a traveling-wave cavity. It is for this reason that we chose to study an HR/AR coated laser, where the AR coating has as low a reflectivity as current technology allows, to maximize the cavity asymmetry. The full mathematical treatment of mirrors with non-unity reflectivity is complicated by the fact that the spatial modes $\Upsilon_m(z)$ are no longer orthogonal, and also that that the $\Upsilon_m(z)$ and the longitudinal overlap factors $\Gamma$ vary with the pumping. This theory will be presented in future work.


\section{Discussion} \label{sec:Discussion}

In order for a mode to oscillate, it must satisfy two conditions: 1) the roundtrip gain must equal the loss, and 2) the roundtrip phase must equal a multiple of $2\pi$. Our theory in Sec.\ \ref{sec:Theory} has treated only the gain condition. The same approach was taken in the description of the original RNGH instability \cite{Risken1968a, Graham1968,Hendow1982a, Hendow1982b, Hillman1982}; the underlying assumption is that the cavity modes are densely spaced, so that a pair of sidebands that satisfies the instability condition for the gain will always be ``close enough" to two cavity modes that satisfy the phase condition. However, the experimental and theoretical developments of the last decade concerning optical parametric oscillation in externally pumped microresonators have shown that the phase condition has a large effect on the oscillation threshold and sideband spacing \cite{Chembo2010}. In microresonator experiments, the detuning between the external pump frequency and the center frequency of the cold cavity mode is a degree of of freedom that must be precisely controlled to achieve the lowest possible instability threshold. In a laser this detuning can not be experimentally controlled, but it most likely varies with the pumping in a deterministic manner and should be properly accounted for in a more complete theory. Furthermore, a parameter that has no analogy in microresonators is $\Delta$, the detuning between the lasing mode $\omega_0$ and the two-level resonance $\omega_{ba}$, which also varies with the pumping and is difficult to control in experiments. To precisely predict the instability threshold would require knowledge of both of these detunings, as well as the GVD.

At this stage, the simplest and most important application of the theory is to help determine whether the observed sidebands are parametrically enhanced or suppressed. Because the theory assumes end mirrors with unity reflectivity, we can only expect Eq.\ \ref{eq:standingwave_sidebandgain} to apply reasonably well to the uncoated QCLs. For each device, $\gamma_D$ is calculated using the theoretical value of $T_{\rm up}$ (calculated from the bandstructure) and the diffusion constant $D = 77$ cm$^2$/s \cite{Faist2013}, giving $\gamma_D = 0.4$ (DS-3.8), 0.49 (TL-4.6), and 0.93 (LL-9.8). For these large values of $\gamma_D$ the PG is strong, and we find from numerically solving Eq.\ \ref{eq:standingwave_sidebandgain} that the FM instability has a lower threshold than the AM instability, regardless of the value of $T_1$. In Appendix \ref{app:sec:numerical}, we show that the theory predicts sideband spacings $\dwsb$ that are consistent with the experimental observations, but underestimates the instability threshold $p_{\rm sb}$. We attribute this discrepancy to the aforementioned detunings and GVD, as well as current inhomogeneity, that our theory neglects.

A more direct method to discriminate between the parametric suppression and enhancement regimes is to compare the observed sideband spacing $\dw_{\rm sb}$ to the crossing frequency $\dw_{\rm cr}$. If $\dw_{\rm sb} < \dw_{\rm cr}$, the sidebands are parametrically suppressed and therefore the FM instability has the lower threshold. Thus, we reason that the sidebands must be FM because the AM state would be an unstable equilibrium. Similarly, if $\dw_{\rm sb} >\dw_{\rm cr}$, the sidebands are parametrically enhanced so for them to be stable they must be AM. Notably, this reasoning depends only on the behavior of $\chi^{(3)}$ as a function of $\dw$; it can therefore be applied to to the HR/AR as well as the uncoated lasers because we do not need to understand the specifics of the PG. To calculate $\dw_{\rm cr}$ from Eq.\ \ref{eq:dw_cr}, we use our measured values of $T_2$ but still need an estimate for the gain recovery time $T_1$. Pump-probe experiments \cite{Choi2008,Choi2009} and theory \cite{Talukder2011} have shown that $T_1$ is around 2\,ps. From Eq.\ \ref{eq:dw_cr}, we see that $\dw_{\rm cr}$ decreases with increasing $T_1$, so if we take $T_1 = 3$\,ps as a generous upper bound on the gain recovery time, we establish a lower bound of $\dw_{\rm cr}$ at 2270\,GHz (DS-3.8), 1730\,GHz (TL-4.6), and 1660\,GHz (LL-9.8). The measured values of $\dwsb$ for each uncoated laser--977\,GHz (DS-3.8), 1259\,GHz (TL-4.6), and 642\,GHz (LL-9.8)--are all substantially smaller than the lower bound on $\dw_{\rm cr}$. This is consistent with the prediction that the uncoated lasers have a lower FM instability threshold than AM threshold, and with these two results we are reasonably confident that the uncoated lasers emit parametrically suppressed FM sidebands. In stark contrast, TL-4.6:HR/AR exhibits a large sideband separation of $\dw_{\rm sb} =2216$\,GHz. If we use the accepted value of $T_1$ equal to 2\,ps, we find $\dw_{\rm cr}= 2120$\,GHz. The observed sideband spacing is slightly larger than $\dw_{\rm cr}$, suggesting the enhancement regime. While a smaller gain recovery time or non-perturbative calculation would raise $\dw_{\rm cr}$ slightly, this is our first hint that TL-4.6:HR/AR emits parametrically enhanced AM sidebands.

The difference in the range of intracavity power over which the harmonic state persists in the uncoated vs. coated lasers, as shown in Fig.\ \ref{fig:LI}, is additional evidence that the uncoated devices operate in the suppression regime and the HR/AR device operates in the enhancement regime. Here we propose a qualitative explanation of this feature. Consider a laser operating in the suppression regime. While the FM state is more stable than the AM state in this regime because it minimizes the amount of gain suppression, the FM state still pays a gain penalty by skipping over modes which would have a greater Lorentzian gain contribution. The dense state could therefore extract more gain from the inverted population, provided that the many modes are phased in such a way that minimizes the amplitude modulation throughout the cavity, thereby avoiding the gain suppression associated with such a low-frequency one-FSR beat note. Indeed, the measurement of negative differential resistance shown in Fig.\,\ref{fig:IVhysteresis} shows that the dense state extracts more gain than the harmonic state. Furthermore, previous experiments have shown that the dense state emission is largely FM, not AM \cite{Hugi2012}. This could explain why the uncoated lasers only exhibit the harmonic state over a small range of current: the laser soon finds a way to transition from the parametrically suppressed FM state to the favored dense state. The fact that TL-4.6:HR/AR exhibits the harmonic state over a large current range suggests that the harmonic state is more stable than the dense state, which could only be true for a harmonic state in the parametric enhancement regime. At a sufficiently high current, when the spectral span of the harmonic state approaches the gain bandwidth, the dense state finally becomes favored for its ability to lase on adjacent modes and extract more incoherent gain, despite no longer benefitting from the parametric enhancement.

We have argued that the dense state is more favored than the FM harmonic state. If this is the case, why do the uncoated QCLs choose to emit a harmonic state at all, and not simply jump from the single-mode state to the dense state as the current is increased? In fact, the spectral hysteresis shown in Fig.\ \ref{fig:spectra_uncoated_rampdown} proves that the dense state {\em is} the favored lasing state down to barely above threshold. However, this state can only be reached by decreasing the current after the laser has already entered the dense state at high current. When the laser starts in a single-mode state and the current is increased, there is clearly a barrier that prevents the transition to the dense state. In general, introducing noise allows a system to overcome energy barriers and explore a larger volume of its state space. It is likely that delayed optical feedback serves as such a noise source, and explains why it is difficult to observe the harmonic state when optical feedback is not eliminated.

\section{Conclusion} \label{sec:Conclusion}
We have experimentally identified the single-mode instability of QCLs, which is characterized by the appearance of sidebands at FP modes not adjacent to the primary lasing mode.  We have seen the behavior in QCLs at three different wavelengths, each based on a different active region design, and with both positive and negative GVD. Therefore, the phenomenon is a general feature of the electron-light dynamics of QCLs. The instability is reached due to the combined contributions of an incoherent gain due to the spatial population grating, and a coherent parametric gain due to the temporal population pulsations. Our theory predicts both an FM instability in situations where the incoherent gain contribution is large, and an AM instability when the incoherent gain contribution is small. This theory extends the RNGH instability of traveling-wave lasers to standing-wave laser cavities. To explore the possibility of AM emission, we coated the QCL facets with an HR and an AR coating to reduce the incoherent gain contribution; indeed, this modification substantially increases the sideband spacing, and it is likely that the waveform is AM. Following the first appearance of sidebands at the instability threshold, our measurements show that increasing the pumping generates more sidebands which preserve the initial spacing. This suggests that a cw QCL can self-start into a phase-locked harmonic frequency comb. We have also placed our observations and theory within historical context, explaining the relation to optically pumped microresonators and the single-mode instability in traveling-wave lasers.

The future direction of this work is clear. At first, we can take guidance from the well-established understanding of microresonators and exploit their similarity with QCLs to further our understanding. The calculation of the instability threshold will be extended to account for GVD, so that the cold-cavity modes are not necessarily equidistant. We must also better understand the nature of the single-mode solution; specifically, how does its detuning from the resonant frequency $\omega_{ba}$, and also its detuning from the cold-cavity mode that it occupies, affect the nature of the instability threshold? We must account quantitatively for the non-unity facet reflections and the precise shape of the mode profile within the cavity. Experimentally, second-order autocorrelation experiments are needed to establish the temporal nature of these short-period waveforms. 


\appendix

\section{Comments on Power vs. Current curves}
\label{app:sec:intracavitypower} 

In the single-mode regime, the intracavity intensity of the single-mode determines the strength of the parametric interaction with the sideband fluctuations. Therefore, we would like to calculate the intracavity intensity from the measured output power. In the distributed loss approximation, the output power is given by
\begin{equation}
P_{out} = \frac{\alpha_m \langle E^2 \rangle L w h}{\sqrt{\mu/\epsilon}}
\end{equation}
where $\alpha_m = \ln[1/(R_1 R_2)]/(2L)$, the length, width, and height of the cavity are $L$, $w$, and $h$, and the time-averaged intensity of the single-mode is $\langle E^2 \rangle = 2 |\curlyE_0|^2$. We are assuming a uniform field intensity in the transverse dimensions, and therefore not worrying about the transverse overlap factor. We can rearrange this equation for the intracavity intensity
\begin{equation}
|\curlyEtilde_0|^2 \equiv \kappa^2 T_1 T_2 |\curlyE_0|^2 = \frac{2 d^2 T_1 T_2 \sqrt{\mu_0/\epsilon_0}}{\hbar^2 n_{\rm eff} \alpha_m L w h} P_{out}. \label{eq:E0Pout}
\end{equation}
With this equation, we can convert the measured total output power of each laser into the intracavity intensity, using our measured values of the refractive index $n_{\rm eff}$ and the dephasing time $T_2$, our best estimates for $d$ and $T_1$, and in the case of the HR/AR laser we have used $R_1=1$, $R_2=0.01$. The result is plotted as a function of $J/J_{\rm th}$ in the inset of Fig.\ \ref{fig:LI}.

\begin{figure}
\includegraphics[scale=0.97]{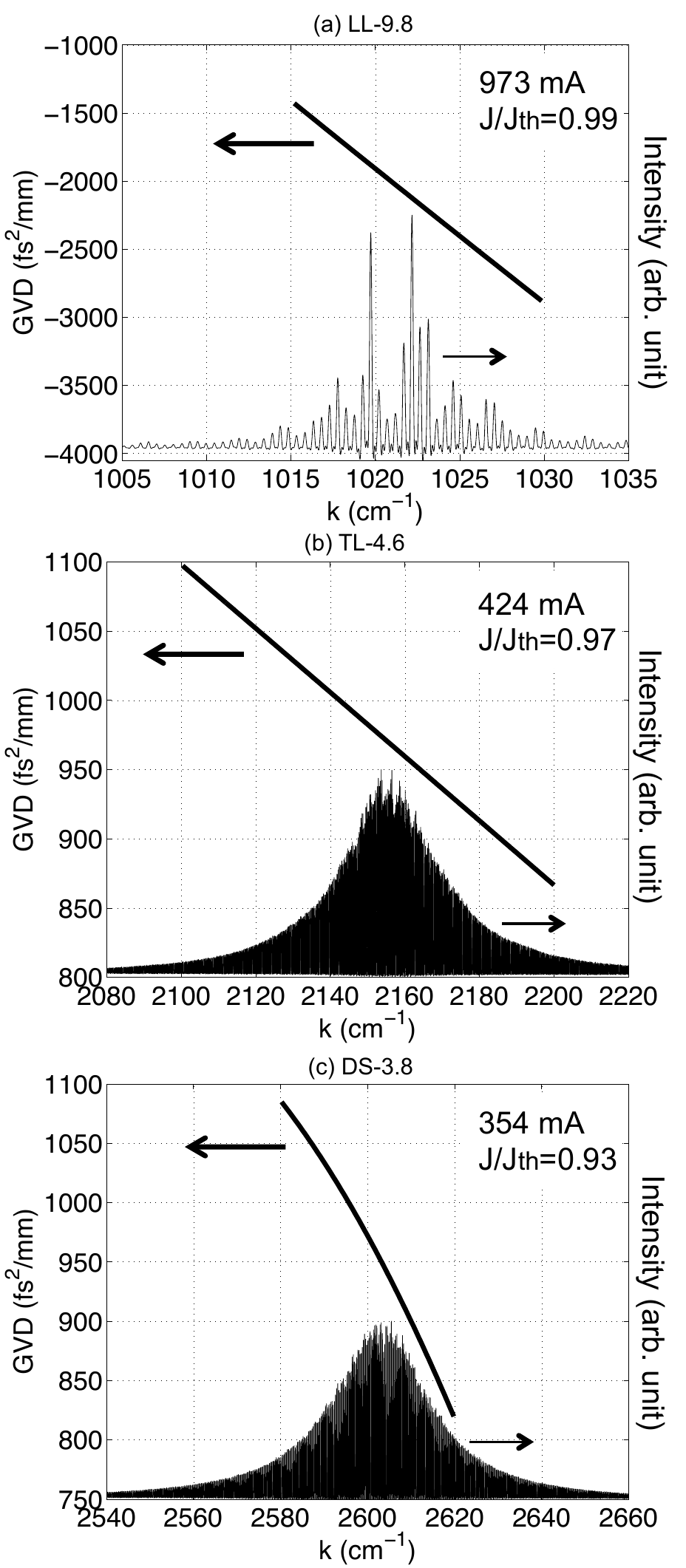}
\caption{\label{fig:GVD}The GVD is extracted from the subthreshold spectrum for (a) LL-9.8 at 973 mA, (b) TL-4.6 at 424 mA, and (c) DS-3.8 at 354 mA. The subthreshold spectra are plotted for reference on an arbitrary linear scale. }
\end{figure}

The theoretical formula for the intracavity intensity is
\begin{equation}
|\curlyEtilde_0|^2 = \frac{p-1}{1+\gamma_D/2},
\end{equation}
where $p \equiv w_{eq}/w_{th}$ is the pump parameter. We emphasize that $p$ is not the same as $J/J_{\rm th}$. The slope of $|\curlyEtilde_0|^2$ vs. $p$ is always between $2/3$ and $1$, depending on the diffusion parameter $\gamma_D$. The reference line in the inset of Fig.\ \ref{fig:LI} is drawn with a slope of one to indicate that each of the $|\curlyEtilde_0|^2$ vs. $J/J_{\rm th}$ curves has a slope greater than one. Therefore, we conclude that $J/J_{\rm th}$ must underestimate $p$. One factor that contributes to this underestimation is the transparency current $J_{\rm trans}$: a fixed amount of current that must be delivered to the active region simply to raise the inversion from a negative number to zero. To understand this simply, suppose that the equilibrium inversion scales like $w_{eq} \propto J - J_{\rm trans}$, and that $J_{\rm trans}$ remains a constant number at threshold and above. Then the pump parameter $p \equiv w_{eq}/w_{th}$ is expressed in terms of $J$ as
\begin{equation}
p = \frac{J - J_{\rm trans}}{J_{\rm th} - J_{\rm trans}}.
\end{equation}
For example, suppose that for a laser with $J_{\rm th} = 500$ mA the harmonic state kicks in at $550$ mA, or $J/J_{\rm th} = 1.1$. If the transparency current was $J_{\rm trans} = 250$ mA,  (in other words, half of the threshold current, which is reasonable for room-temperature QCLs), then the pump parameter at the harmonic state onset would be $p = (550 - 250)/(500 - 250) = 1.2$. Thus, $J/J_{\rm th}$ underestimates $p$.

A more rigorous study is required to determine $J_{\rm trans}$ for each laser, which can be done by measuring many lasers of the same active region but different lengths. Once $J_{\rm trans}$ is known, the slope of $|\curlyEtilde_0|^2$ vs. $p$ should fall between $2/3$ and 1 and in principle a value for $\gamma_D$ can be extracted, allowing one to quantify the amount of diffusion present.

\section{Group velocity dispersion}
\label{app:sec:GVD}

It has recently been shown that the group velocity dispersion (GVD) of the QCL is an important parameter in determining the spectral properties of the dense state \cite{Villares2016}, and it is reasonable to assume that GVD plays a role in determining some properties of the harmonic state as well. We present the measured GVD of the three uncoated devices in Fig.\,\ref{fig:GVD}. The GVD is extracted from a measurement of the subthreshold amplified spontaneous emission by the method of \cite{Hofstetter1999}. A more sensitive InSb detector (compared to HgCdTe) can be used for the lower wavelength devices TL-4.6 and DS-3.8, allowing the measurement to be performed further below threshold. This yields a broader spontaneous emission spectrum, and a larger bandwidth over which the GVD can be extracted. Devices TL-4.6 and DS-3.8 both have positive GVD, around 950 fs$^2$/mm at the center of their gain spectra, while LL-9.8 has a negative GVD around -2000 s$^2$/mm. The change in sign of the GVD is expected because the zero-GVD point of InP is around 5.5 $\mu$m.

Although the theoretical analysis presented in this work has for the most part neglected GVD, the purpose of this measurement is to demonstrate that the harmonic state exists over a wide range of GVD, including both positive and negative values. We hope that including this data here will help guide future work on the role played by GVD in the properties of the harmonic state.

\section{Beat note in dense state}
\label{app:sec:beatnote}

\begin{figure}
\includegraphics[scale=0.47]{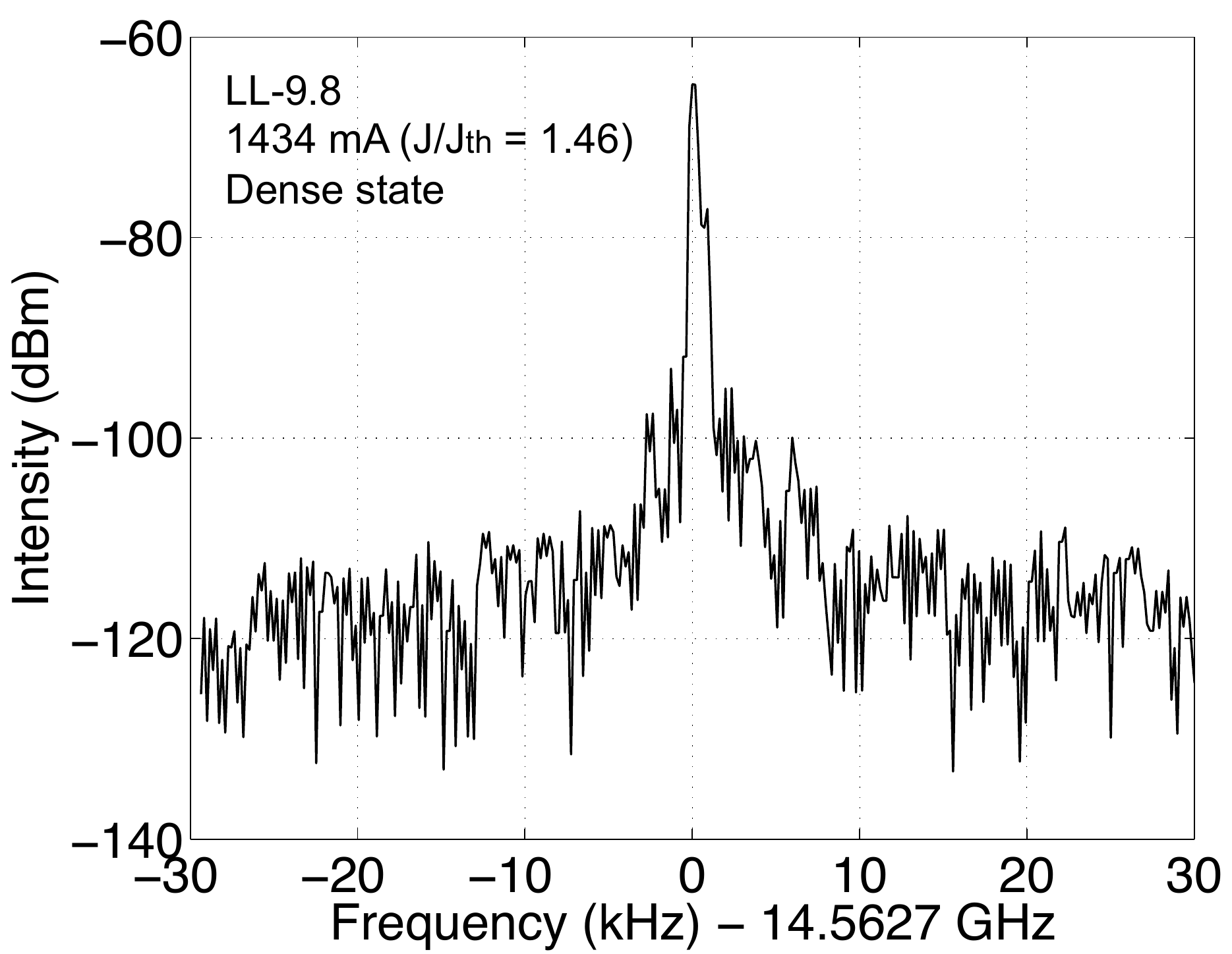}
\caption{\label{fig:beatnote} The beat note of LL-9.8 at 1434 mA, the current at which the dense state appeared. The resolution bandwidth of the spectrum analyzer is 430 Hz.}
\end{figure}

In the dense state the mode spacing is one FSR ($\approx 15$ GHz for a 3-mm long QCL), which is low enough for the beat note to be measured by standard techniques, unlike the case of the harmonic state. We measure the beat note by the electrical technique: the current source is passed from the DC to the AC+DC port of a high-frequency bias tee (Tektronix PSPL5544) en route to the QCL, so that current modulation induced by any intracavity intensity modulation can be measured by a spectrum analyzer connected to the AC port of the bias tee. In Fig.\,\ref{fig:beatnote}, the beat note of LL-9.8 is shown at 1434 mA, the current at which the dense state appeared on this particular upward ramp up of the current. The width of the beat note is on the order of a few kHz. Devices TL-4.6 and DS-3.8 exhibited similar beat notes at the onset of the dense state (not plotted). As the current was increased further, different regimes could be observed such as the appearance of multiple closely-spaced beat notes, or a broadening of the beat note. Because the focus of the current work is on the harmonic state, these diverse behaviors of the dense state will be explored in future work.

\section{Gain recovery time vs. upper state lifetime}
\label{app:sec:TupvsT1}
In a QCL, the upper state lifetime $T_{\rm up}$ tells us how long an electron sits in the upper state before making a nonradiative transition to the lower state. From here, it takes some additional time to travel through the injector region and tunnel into the upper level of the next stage. This additional amount of time is the bottleneck that determines the gain recovery time. The Maxwell-Bloch equations, by making the two-level approximation, cannot account for the full complexity of the QCL, and only provide us with one carrier relaxation time, which we have called $T_1$. This begs the question: does $T_1$ represent the upper state lifetime or the gain recovery time? The answer is that it depends on what you want to calculate. In the steady-state single-mode regime, we find that the output power and population inversion are functions of $T_1$ due to diffusion; here, we argue that $T_1$ should represent the upper state lifetime $T_{\rm up}$, because $T_{\rm up}$ tells us how much time an electron in the upper state has to diffuse before transitioning to the lower state. It is for this reason that $T_{\rm up}$ appears in the definition of $\gamma_D$, $\gamma_D = (1+4 k^2 D T_{\rm up})^{-1}$, rather than $T_1$. In dynamical situations, on the other hand, the intensity of the field varies with time and we are interested in how the population inversion responds. We argue that this response is determined by the gain recovery time, not the upper state lifetime, because the response definitely depends on how long it takes an electron to get from one active stage to the next. To summarize it concisely, the upper state lifetime is used for the calculation of the population grating (PG), but the gain recovery time is used for the calculation of the population pulsations (PPs). Since the bulk of our manuscript deals with PPs, we chose simply to call $T_1$ the gain recovery time, rather than name a new time scale such as $T_{\rm gr}$, for instance. We hope that this does not confuse the reader.

\section{Theory: single-mode solution}
\label{app:sec:singlemodetheory}
This appendix gives a more detailed derivation of the single-mode solution presented in Sec.\,\ref{sec:PG}, including the intracavity power as a function of pumping, and the population inversion as a function of position and pumping.

For a two level system with upper state $|a\rangle$ and lower state $|b\rangle$, the material equations in the non-rotating frame and the field equation are
\begin{align}
\frac{d \rho_{ab}}{dt} &= - i \omega_{ba} \rho_{ab} - \frac{id}{\hbar} E(t) w - \frac{\rho_{ab}}{T_2} \\
\frac{dw}{dt} &= \frac{-2 i d}{\hbar} E(t)(\rho_{ab} - \rho_{ab}^*) + \frac{w_{eq}-w}{T_1} + D \frac{\partial^2 w}{\partial z^2} \\
\frac{\partial^2 E}{\partial z^2} & - \frac{1}{c^2} \frac{\partial^2 E}{\partial t^2} = Nd \mu \frac{\partial^2}{\partial t^2} (\rho_{ab} + \rho_{ab}^*).
\end{align}
We emphasize that these equations are in the non-rotating frame, whereas the equations we have used in the main text \cite{Allen1987} were already in the rotating frame and the RWA had already been applied. However, since we are here dealing with two counter-propagating waves, we chose to more closely follow the approach in \cite{Gordon2008}. We make the following ansatzes:
\begin{align}
E(z,t) &= \frac{1}{\sqrt{2}} \left [ \curlyE_R(z,t) e^{-i(\omega t - k z)} + \curlyE_L(z,t) e^{-i(\omega t + k z)} +c.c. \right] \\
\rho_{ab} (z,t) &= \eta_R^*(z,t) e^{-i(\omega t - k z)} + \eta_L^* (z,t) e^{-i(\omega t + k z)} \\
w(z,t) &=w_{\rm DC}(z,t) + w_2 (z,t) e^{i2kz} + w_2^*(z,t) e^{-i2kz}.
\end{align}
(We use the subscript ``DC" rather than ``0" for the spatial average of the population inversion, $w_{\rm DC}$, because the subscript 0 is used throughout the text to refer to the primary mode. No such ambiguity occurs for the subscript ``2.") Plugging the ansatzes into the differential equations, and making the RWA as well as the slowly-varying envelope approximation (SVEA) yields the following equations:
\begin{align}
\frac{d \eta_R^*}{dt} &= \frac{-i \kappa }{2\sqrt{2}} (\curlyE_R w_{\rm DC} + \curlyE_L w_2) - \left( \frac{1}{T_2} + i \Delta \right) \eta^*_R \\
\frac{d \eta_L^*}{dt} &= \frac{-i \kappa}{2\sqrt{2}} (\curlyE_L w_{\rm DC} + \curlyE_R w_2^*) - \left( \frac{1}{T_2} + i \Delta \right) \eta^*_L\\
\frac{d w_{\rm DC}}{dt} &= \frac{i\kappa}{\sqrt{2}} (\curlyE_R \eta_R + \curlyE_L \eta_L - c.c.) + \frac{w_{eq} - w_{\rm DC}}{T_1} \\
\frac{d w_2}{dt} &= \frac{i \kappa}{\sqrt{2}} (\curlyE_R \eta_L - \curlyE_L^* \eta_R^* ) - \frac{w_2}{T_1} - 4k^2 D w_2 \\
\frac{1}{c} \frac{\partial \curlyE_R}{\partial t} &= - \frac{\partial \curlyE_R}{\partial z} + \frac{i \sqrt{2} \alpha}{\kappa T_2} \eta_R^* - \frac{\ell_0}{2} \curlyE_R \\
\frac{1}{c} \frac{\partial \curlyE_L}{\partial t} &= + \frac{\partial \curlyE_L}{\partial z} + \frac{i \sqrt{2} \alpha}{\kappa T_2} \eta_L^* - \frac{\ell_0}{2} \curlyE_L 
\end{align}
where $\kappa = 2d/\hbar$, $\alpha= N \omega T_2 d^2 \sqrt{\mu/\epsilon}/\hbar$ is the Beer coefficient of the material, and $\Delta = \omega_{ba} - \omega$ is the detuning of the field from the atomic resonance frequency. The loss term $\ell_0$ has been added to the field equation heuristically, and in this context it represents only the waveguide loss.

We solve for the single-mode solution by setting the time-derivatives to zero and the slowly-varying envelope functions to be constants. In doing so, we are now making the distributed loss approximation because we are not allowing the fields to grow in space. Thus, $\ell_0$ must now be taken to be the total loss, waveguide plus mirror loss, which we call $\ell$. (In the main text, the loss is expressed by the rate $\bar{\ell}$, which is simply given by $\bar{\ell} \equiv \ell c$, and the Beer coefficient $\alpha$ is similarly converted into a rate $\bar{\alpha}\equiv \alpha c$.) We take $\Delta=0$ for simplicity, because the single-mode will lase very close to the peak of the gain spectrum. We denote the steady-state field amplitudes by $\curlyE_R=\curlyE_L = \curlyE_0$ and find the LI curve
\begin{equation}
|\curlyEtilde_0|^2  = \frac{p-1}{1+\gamma_D/2} \label{app:eq:LI}
\end{equation}
where $\curlyEtilde_0 \equiv \kappa \sqrt{T_1 T_2} \curlyE_0$, $p=w_{eq}/w_{th}$, $w_{th} = \bar{\ell}/\bar{\alpha}$, and $\gamma_D = (1+4 k^2 D T_1)^{-1}$ is the diffusion parameter. Based on the discussion in Appendix \ref{app:sec:TupvsT1}, however, we know $T_1$ represents the upper state lifetime $T_{\rm up}$, so we redefine $\gamma_D = (1+4 k^2 D T_{\rm up})^{-1}$. The steady-state population $w_0(z)$ is given by 
\begin{equation}
w_0(z) = w_{th} \left[ 1 + \frac{\gamma_D}{2} \frac{p-1}{1 + \gamma_D/2} - \gamma_D \frac{p-1}{1+\gamma_D/2} \cos(2k_0 z) \right] \label{app:eq:w0(z)_standingwave}.
\end{equation}
As the diffusivity or $T_{\rm up}$ increases, causing $\gamma_D$ to approach zero, the population grating is ``washed out'' and $w_0(z)$ is uniformly equal to $w_{th}$.

We have made the approximation that the pumping $p$ is uniform in space. For an electrically injected QCL, this is equivalent to assuming that the injected current density $J$ is uniform throughout the cavity. In fact, however, the resistance of the active region is lower in the field antinodes because stimulated emission increases the rate of electron transport. Therefore, assuming a constant voltage across the active region, more current will flow through the lower-resistance antinodes, an effect which reduces the amplitude of the population grating even in the absence of any lateral carrier diffusion. It helps to picture the active region as two resistors in series, one constant ``background'' resistance in series with one whose resistance drops with increasing light intensity. Devices with a lower background resistance will be prone to greater current inhomogeneity. The magnitude of this effect can be estimated from the kink in the current-voltage curve above threshold, which shows how much the photon field reduces the device resistance, but we have ignored this effect in our current work. One consequence of ignoring this inhomogeneity of the current is that we overestimate the amplitude of the population grating. This likely contributes to our theory's underestimation of the instability threshold $p_{\rm sb}$, as discussed in Appendix \ref{app:sec:numerical}.

\section{Theory: population pulsations}
\label{app:sec:PPs}
This appendix gives a more detailed derivation of the population pulsations presented in Sec.\,\ref{sec:PP}, and demonstrates how to include nonzero detuning $\Delta$ and GVD into the formalism.

We begin by imagining a small volume of dipoles subject to a spatially uniform $E$-field to develop an understanding of the non-linear effects caused by the Bloch dynamics. The electric field is given by
\begin{equation}
E(t) = \curlyE(t) e^{i\omega t} + c.c.
\end{equation}
The Bloch equations in the rotating wave approximation are
\begin{align}
\dot{\sigma} &= \left( i \Delta - \frac{1}{T_2} \right) \sigma + \frac{i \kappa}{2} w \curlyE \label{app:eq:Bloch_sigma} \\
\dot{w} &= i \kappa (\curlyE^* \sigma - \curlyE \sigma^*) - \frac{w - w_{eq}}{T_1} \label{app:eq:Bloch_w}
\end{align}
where $\sigma$ is the off-diagonal element of the density matrix in the rotating frame, $w$ is the population inversion (positive when inverted), $\Delta = \omega_{ba} - \omega$ is the detuning between the applied field and the resonant frequency of the two-level system, $T_1$ is the (longitudinal) population relaxation time, $T_2$ is the (transverse) dephasing time, $\kappa \equiv 2d/\hbar$ is the coupling constant where $d$ is the dipole matrix element (assumed to be real) and $\hbar$ is Planck's constant, and $w_{eq}$ is the equilibrium population inversion in the absence of any electric field which is determined by the pumping. (Note that these equations are identical to Eqs. 3.19(a)-(c) in \cite{Allen1987}, except that we have allowed $\curlyE$ to be complex and left the off-diagonal component of the density matrix in complex notation rather than writing $\sigma = (u + i v)/2$.) With these conventions, the macroscopic polarization $P$ (dipole moment per volume) in a region with  a volume density of $N$ dipoles is given by
\begin{equation}
P(t)=Nd \sigma e^{i\omega t} + c.c.
\end{equation}

First, we consider the effect of a monochromatic field at frequency $\omega$, obtained from Eqs.\ \ref{app:eq:Bloch_sigma}-\ref{app:eq:Bloch_w} by setting $\curlyE(t) = \curlyE_0$ and all time derivatives to zero. The result is a steady-state polarization $\sigma_0$ and population inversion $w_0$ given by
\begin{align}
\sigma_0 &= \frac{i \kappa T_2}{2 (1 - i \Delta T_2)} w_0 \curlyE_0 \\
w_0 &= \frac{w_{eq}}{1 + \frac{\kappa^2 T_1 T_2 |\curlyE_0|^2}{1 + (\Delta T_2)^2}} 
\end{align}
Note that the population inversion $w_0$ is saturated as the field strength $\curlyE_0$ increases: this is responsible for saturable loss (when $w_{eq}<0$) and saturable gain (when $w_{eq}>0$).

N.B. In our equations so far, we have said the frequency of the field is $\omega$. Later on, we refer to the primary mode frequency as $\omega_0$. For our purposes here, $\omega$ and $\omega_0$ are interchangeable. In future work, this will not be the case. In analogy with the theory developed for microresonators \cite{Chembo2010a}, we plan in future work to adopt the convention that $\omega_0$ represents the center-frequency of the cold-cavity mode that is lasing. However, the lasing frequency $\omega$ can be detuned from this cold-cavity resonance due to small frequency-pulling effects when one accounts for the hot cavity. This detuning is an important parameter in microresonators, where the pump frequency $\omega$ can be controllably tuned away from $\omega_0$, allowing one to compensate for GVD and optimize comb generation. We have not accounted for such a detuning in our work, and therefore $\omega$ and $\omega_0$ are interchangeable.

\subsection{Two-frequency operation}

Next, we consider the $E$-field
\begin{equation}
\curlyE = \curlyE_0 + \curlyE_+ e^{i \dw t}
\end{equation}
which consists of the strong field $\curlyE_0$ at frequency $\omega$ superposed with the much weaker field $\curlyE_+$ detuned from $\omega$ by $\dw$. A polarization will of course be induced at $\omega+\dw$. However, a polarization at $\omega - \dw$ also results due to the beat note at $\dw$ which modulates the intensity: the resulting modulation of the population inversion with time (i.e., a population pulsation) leads to nonlinear frequency mixing. We express the full polarization as
\begin{equation}
P(t) = \sum_{m=-,0,+} \curlyP_m e^{i \omega_m t} + c.c.
\end{equation}
where $\omega_+ \equiv \omega + \dw$ and $\omega_- \equiv \omega - \dw$. We can solve for the polarization as done in \cite{Boyd2003}, keeping only terms to first order in the weak field $\curlyE_+$, which gives
\begin{align}
\curlyP_0 &= \frac{i \epsilon}{\omega_{ba}} \bar{\alpha} w_0 \frac{\curlyE_0}{1 - i \Delta T_2} \\
\curlyP_+  &= \frac{i \epsilon}{\omega_{ba}} \bar{\alpha} w_0 \left[ \frac{\curlyE_+}{1 - i (\Delta - \dw) T_2} + \Lambda^+_+ \curlyEtilde_0 \curlyEtilde_0^* \curlyE_+  \right] \\
\curlyP_-  &= \frac{i \epsilon}{\omega_{ba}} \bar{\alpha} w_0  \Lambda^+_- \curlyEtilde_0 \curlyEtilde_0 \curlyE_+^*
\end{align}
where
\begin{widetext}
\begin{align}
\Lambda^+_+ &= \frac{ - (1+i \dw T_2/2)[1 + i(\Delta + \dw)T_2]/[1 - i (\Delta - \dw) T_2] }{(1 + i \Delta T_2) \left[ (1 + i \dw T_1) [1 + i(\Delta + \dw)T_2] [1 - i(\Delta - \dw)T_2] + (1+ i\dw T_2) |\curlyEtilde_0|^2 \right] } \label{app:eq:Lambda++} \\
\Lambda^+_- &= \frac{ - (1-i \dw T_2/2) }{(1 - i \Delta T_2) \left[ (1 - i \dw T_1) [1 + i(\Delta - \dw)T_2] [1 - i(\Delta + \dw)T_2] + (1- i\dw T_2) |\curlyEtilde_0|^2 \right] }  \label{app:eq:Lambda+-} 
\end{align}
\end{widetext}
are the self-mixing and cross-mixing coupling coefficients, respectively. We consider the dipoles to be embedded in a host medium of permittivity $\epsilon$ and permeability $\mu$. (We adopt the convention of \cite{Siegman1986}: $\epsilon$, $\mu$ and the speed of light $c=1/\sqrt{\epsilon \mu}$ always take their values in the background host medium.) Many of the material properties of the two-level system are lumped into the ``Beer rate''
\begin{equation}
\bar{\alpha} = \frac{N d^2 T_2 \omega_{ba} c \sqrt{\mu/\epsilon}}{\hbar},
\end{equation}
which is related to the more familiar Beer absorption coefficient $\alpha$ (with units of inverse length) that appears in BeerÕs law of absorption by $\bar{\alpha} = \alpha c$. (Note, however, that in our expressions for the polarization due to the two-level system, all factors of $\epsilon$ and $\mu$ drop out; that is, these expressions do not contain the polarization contributions due to the background medium.) The central mode amplitude $\curlyE_0$ has been normalized such that $\curlyEtilde_0 \equiv \kappa \sqrt{T_1 T_2} \curlyE_0$. Note that $\curlyP_0$ is unaffected to first order in $\curlyE_+$. The polarization $\curlyP_+$ comes from two contributions. First, there is the linear contribution from the Lorentz oscillator which $\curlyE_+$ would induce even in the absence of the strong field $\curlyE_0$. Second, there is a contribution due to the PP which is described by the term $\Lambda_+^+$. The term $\curlyP_-$ is due solely to the PP and is governed by $\Lambda^+_-$. Note that the full polarization is directly proportional to the steady-state population inversion $w_0$; this will be important when we generalize our results to standing-wave cavities, where $w_0$ varies with position.

Now that we have the polarization, we can calculate the gain seen by the sideband field. We define the gain $\bar{g}$ (with dimension of frequency) of the sideband as the power density generated at $\omega + \dw$ by the interaction of the field with the dipoles--considering only field and polarization terms oscillating at $\omega + \dw$--divided by the energy density of the exciting sideband field, or
\begin{align}
\bar{g}_+ & \equiv - \frac{\langle E \dot{P} \rangle_+} {2 \epsilon |\curlyE_+|^2} \\
&= \frac{i\omega_+(\curlyE_+ \curlyP_+^* - \curlyE_+^* \curlyP_+) }{2 \epsilon |\curlyE_+|^2}
\end{align}

\subsubsection{$\Delta=0$}
Here we consider the case of zero detuning, $\Delta=0$, which simplifies the mathematical expressions considerably. Under this scenario, we denote the self-mixing coefficient $\Lambda^+_+$ by $\Lambda$, where
\begin{equation}
\Lambda = \frac{ - (1+i \dw T_2/2) }{\left[ (1 + i \dw T_1) (1 + i \dw T_2)^2  + (1+ i\dw T_2) |\curlyEtilde_0|^2 \right] } \label{app:eq:Lambda++_Delta=0},
\end{equation}
and it is simple to show that the cross-coupling coefficient $\Lambda^+_-$ is $\Lambda^*$. The gain of the sideband field is found to be
\begin{equation}
\bar{g}_+ = \bar{\alpha} w_0 \left[ \frac{1}{1+ (\dw T_2)^2} + {\rm Real} ( \Lambda) |\curlyEtilde_0|^2  \right].
\end{equation}
(We have used $(\omega+\dw)/\omega_{ba} \approx 1$.) Thus, the gain can be nicely divided up into a contribution from the Lorentz oscillator and a contribution from the PP. All of this is proportional to $\bar{\alpha} w_0$: $\bar{\alpha}$ gives you the gain of a weak field tuned to line-center in a perfectly inverted medium (or alternatively, the loss seen by a weak field tuned to line-center in a material in its ground state),  and $w_0$ gives you the expectation value of finding an electron in the excited state (equal to 1 when excited, -1 when in the ground state, and 0 at transparency). Note that ${\rm Real} (\Lambda)$ can be positive or negative, which we will discuss shortly.

\subsection{Three-frequency operation}

Of course, the polarization created at $\omega - \dw$ will create a field at that frequency, which is precisely why in the experiments we always observe the two sidebands appearing simultaneously. One sideband cannot exist in isolation when the mixing terms naturally couple them together. Therefore, we need to consider the field
\begin{equation}
\curlyE = \curlyE_0 + \curlyE_+ e^{i \dw t} + \curlyE_- e^{-i \dw t}.
\end{equation}
The polarization at each sideband frequency now contains a Lorentzian term, a self-mixing term, and a cross-mixing term:
\begin{widetext}
\begin{align}
\curlyP_+ &= \frac{i \epsilon}{\omega_{ba}} \bar{\alpha} w_0 \left[ \frac{\curlyE_+}{[1 - i (\Delta - \dw) T_2]} + \Lambda^+_+ \curlyEtilde_0 \curlyEtilde_0^*  \curlyE_+ + \Lambda^-_+ \curlyEtilde_0 \curlyEtilde_0 \curlyE_-^*\right] \label{app:eq:P+}\\
\curlyP_- &= \frac{i \epsilon}{\omega_{ba}} \bar{\alpha} w_0 \left[ \frac{\curlyE_-}{[1 - i (\Delta + \dw) T_2]} + \Lambda^-_- \curlyEtilde_0 \curlyEtilde_0^*  \curlyE_- + \Lambda^+_- \curlyEtilde_0 \curlyEtilde_0 \curlyE_+^*\right] \label{app:eq:P-}
\end{align}
\end{widetext}
where $\Lambda^-_-$ and $\Lambda^-_+$ are obtained by making the substitution $\dw \rightarrow - \dw$ in the expressions for $\Lambda^+_+$ and $\Lambda^+_-$, respectively, given in Eqs.\  \ref{app:eq:Lambda++}-\ref{app:eq:Lambda+-} .

\subsubsection{$\Delta=0$}
Let us again focus on the case $\Delta=0$, for which the polarization at each sideband simplifies to
\begin{align}
\curlyP_+ &= \frac{i \epsilon}{\omega_{ba}} \bar{\alpha} w_0 \left[ \frac{\curlyE_+}{[1 + i  \dw T_2]} + \Lambda \curlyEtilde_0 \curlyEtilde_0^*  \curlyE_+ + \Lambda \curlyEtilde_0 \curlyEtilde_0 \curlyE_-^*\right] \\
\curlyP_- &= \frac{i \epsilon}{\omega_{ba}} \bar{\alpha} w_0 \left[ \frac{\curlyE_-}{[1 - i  \dw T_2]} + \Lambda^* \curlyEtilde_0 \curlyEtilde_0^*  \curlyE_- + \Lambda^* \curlyEtilde_0 \curlyEtilde_0 \curlyE_+^*\right] ,
\end{align}
where $\Lambda$ is simply $\Lambda^+_+$ evaluated for $\Delta=0$. We see the nice property that when $\Delta=0$, $\Lambda^+_+ = \Lambda^-_+$ ($\equiv \Lambda$), and  $\Lambda^-_- = \Lambda^+_-$ ($\equiv \Lambda^*$); in other words, the self- and cross-mixing coupling coefficients are equal.

The gain $\bar{g}_+$ of the positive sideband is
\begin{equation}
\bar{g}_+ = \bar{\alpha} w_0 \left\{ \frac{1}{1+ (\dw T_2)^2} + {\rm Real} \left[ \Lambda |\curlyEtilde_0|^2 \left( 1 + \frac{\curlyEtilde_0^2 \curlyE_-^*}{|\curlyEtilde_0|^2 \curlyE_+} \right) \right] \right\},\label{app:eq:g+}
\end{equation}
and a similar expression holds for the minus sideband. This equation tells us that the PP contribution to the gain depends on the phase and amplitude relationships of $\curlyE_0$, $\curlyE_-$, and $\curlyE_+$, which is not too surprising because the amplitude of the PP itself is sensitive to these parameters. Without loss of generality, we can take $\curlyE_0$ to be real. If $\curlyE_+ = \curlyE_-^*$, then the two sidebands' contributions to the beat note at $\dw$ add constructively, resulting in a field whose amplitude modulation (AM) is twice the strength of a field with only one sideband. If $\curlyE_+ = -\curlyE_-^*$, then the two sidebands' contributions to the beat note at $\dw$ destructively cancel and there is no longer any amplitude modulation at frequency $\dw$. We refer to such a field as frequency-modulated (FM). We see from Eq.\ \ref{app:eq:g+} that the AM sidebands therefore experience a PP contribution to the gain that is twice as large as the single sideband case, while the FM sidebands experiences only the background Lorentzian gain, consistent with the fact that there is no PP in this case. We summarize this with the formula for the gain $\bar{g}$ of each sideband for the case of equal-amplitude sidebands ($|\curlyE_+|=|\curlyE_-|$),
\begin{equation}
\bar{g} = \bar{\alpha} w_0 \left[ \frac{1}{1+ (\dw T_2)^2} + {\rm Real}(\Lambda)|\curlyEtilde_0|^2 \times \left\{
\begin{array}{cc}
2  & {\rm ;\ AM} \\
0 & {\rm ;\ FM}
\end{array}
\right.
\right].
\end{equation}
Note that for a superposition of AM and FM, the gain due to the PP will fall between 0 and 2 times the factor ${\rm Real} (\Lambda)|\curlyEtilde_0|^2$.

\section{Theory: instability threshold}
\label{app:sec:instabilitythreshold}

This appendix gives a more detailed derivation of the instability threshold presented in Sec.\,\ref{sec:instabilitythreshold}, and demonstrates that the gain seen by the sidebands is due to a contribution from the population grating and another from the population pulsations.

When a continuous-wave (cw) laser is pumped at its lasing threshold, only a single frequency of light--the one nearest the gain peak that also satisfies the roundtrip phase condition--has sufficient gain to overcome the roundtrip loss and begins to lase. As the pumping is increased, the single-mode solution yields to multimode operation; this is known as the single-mode instability. Our goal is to determine 1) how hard to pump the laser to reach the single-mode instability and 2) which new frequencies start lasing.

Consider a laser pumped above threshold that is lasing on a single-mode, which we refer to as the primary or central mode. If another mode is to lase, it must be seeded by a spontaneously generated photon at a different frequency. This photon will necessarily create a beat note through its coexistence with the primary mode, resulting in a population pulsation. The gain seen by the new frequency must therefore account for this parametric gain in addition to the background Lorentzian gain. Furthermore, the PP couples the sideband to the symmetrically detuned sideband frequency on the other side of the primary mode, so we should in general assume the presence of both sidebands. Because the instability threshold depends on the cavity geometry, we will consider a traveling-wave laser as well as a standing-wave laser. In both cases, the strategy is the same. First, we solve for the single-mode intensity $\curlyE_0$ and the population inversion $w_0(z)$ as a function of the pumping, entirely neglecting the sidebands. Knowing this, we can then calculate the sideband gain in the presence of the primary mode.

We start with the wave equation
\begin{equation}
\frac{\partial^2 E}{\partial z^2} - \frac{1}{c^2} \frac{\partial^2 E}{\partial t^2} = \mu \frac{\partial^2 P}{\partial t^2}.
\end{equation}
Following the approach used to calculate the optical parametric oscillation threshold in optically pumped microresonators \cite{Chembo2010a}, we expand the field in terms of the cold cavity modes,
\begin{equation}
E(z,t) = \sum_{m=-,0,+} \curlyE_m(t) \Upsilon_m(z) e^{i \omega_m t} + c.c.
\end{equation}
The spatial modes obey the normalization condition
\begin{equation}
\frac{1}{L} \int_0^L dz\ |\Upsilon_m(z)|^2 = 1.
\end{equation}
When group velocity dispersion (GVD) is non-zero, the two modes $\omega_+$ and $\omega_-$ will not be equidistant from $\omega_0$. We have also assumed that the spatial and temporal dependence of the modes can be separated. This is a good approximation in the case of a laser, because we know the intracavity field will be sharply resonant at the modes. The spatial variation of the polarization can be described by making the substitution $\curlyE_m \rightarrow \curlyE_m \Upsilon_m(z)$ and $w_0 \rightarrow w_0(z)$ into the polarization Eqs.\ \ref{app:eq:P+}-\ref{app:eq:P-}, which results in the polarization
\begin{equation}
P(z,t) = \sum_{m=-,0,+} \curlyP_m(z,t) e^{i \omega_m t} + c.c.
\end{equation}
where
\begin{widetext}


\begin{align}
\curlyP_+ (z,t) &= \frac{i \epsilon}{\omega_{ba}} \bar{\alpha} w_0(z) \left[ \frac{\curlyE_+ \Upsilon_+(z)}{[1 - i (\Delta - \dw) T_2]} \right. + \Lambda^+_+(z) |\Upsilon_0(z)|^2 \Upsilon_+ (z) |\curlyEtilde_0|^2   \curlyE_+ + \left. \Lambda^-_+(z)  \Upsilon_0(z)^2 \Upsilon_-^*(z) e^{i \bar{\omega} t} \curlyEtilde_0^2  \curlyE_-^*  \right] \\
\curlyP_- (z,t) &= \frac{i \epsilon}{\omega_{ba}} \bar{\alpha} w_0(z) \left[ \frac{\curlyE_- \Upsilon_-(z)}{[1 - i (\Delta + \dw) T_2]} \right. + \Lambda^-_-(z) |\Upsilon_0(z)|^2 \Upsilon_- (z) |\curlyEtilde_0|^2   \curlyE_- + \left. \Lambda^+_-(z) \Upsilon_0(z)^2 \Upsilon_+^*(z) e^{i \bar{\omega} t} \curlyEtilde_0^2  \curlyE_+^*  \right].
\end{align}
\end{widetext}

We have introduced $\bar{\omega} \equiv 2 \omega_0 - \omega_+ - \omega_-$, the deviation of the cold cavity modes from equal spacing. Note that the $\Lambda$s now depend on $z$ due to the term in their denominators dependent on the primary mode amplitude. Because we no longer demand that the two sidebands have the same detuning $\dw$, $\Lambda^+_+$ and $\Lambda^+_-$ should, strictly speaking, be calculated using the detuning $\dw_+ = \omega_+ - \omega_0$, while $\Lambda^-_-$ and $\Lambda^-_+$ should depend on $\dw_- = \omega_0 - \omega_-$. In practice, we can ignore this difference in the $\Lambda$s; the term $e^{i \bar{\omega} t}$ captures the most important effect of GVD.

Plugging everything into the wave equation gives
\begin{widetext}
\begin{equation}
\sum_m \left( \frac{d^2 \Upsilon_m}{dz^2} + \frac{\omega_m^2}{c^2} \Upsilon_m \right) \curlyE_m e^{i \omega_m t} - \frac{2i}{c^2} \sum_m \omega_m \frac{d \curlyE_m}{dt} \Upsilon_m e^{i \omega_m t} = \mu \sum_m -\omega_m^2 ( \curlyP_m - \curlyP_{m,{\rm loss}})e^{i \omega_m t} \label{app:eq:coupledmodes}
\end{equation}
\end{widetext}
where the slowly-varying-envelope approximation allowed us to ignore second time derivatives of $\curlyE_m$ on the left-hand side, and first and second derivatives of $\curlyE_m$ on the right-hand side. The spatial modes $\Upsilon_m(z)$ are chosen so that the first term on the LHS equals zero. The loss of each mode has been added to the equation in the form of a polarization contribution; we assume each mode has the same linear loss, which can be expressed
\begin{equation}
\curlyP_{m,{\rm loss}}(z,t) = \frac{i \epsilon}{\omega_{ba}} \bar{\ell} \Upsilon_m(z) \curlyE_m(t).
\end{equation}
Equation \ref{app:eq:coupledmodes} couples all of the modes $\curlyE_m$. We can project this equation onto each mode by multiplying by $\Upsilon_n (z)$ and integrating over the length of the laser cavity, thus taking advantage of the orthonormality of the spatial modes $\Upsilon_m(z)$, and then equating terms which oscillate at the same frequency (since terms with different frequencies will not affect the time-averaged gain seen by a mode). The result is one equation for the central mode
\begin{equation}
\dot{\curlyE}_0 = \left[ -\frac{\bar{\ell}}{2} + \frac{\bar{\alpha}}{2(1 - i \Delta T_2)} \int \frac{dz}{L} \ w_0 (z) |\Upsilon_0(z)|^2 \right] \curlyE_0, \label{app:eq:dE0dt_general}
\end{equation} 
and one equation for each of the sidebands,
\begin{widetext}
\begin{align}
\dot{\curlyE}_+ = -\frac{\bar{\ell}}{2} \curlyE_+ + \frac{\bar{\alpha}}{2} & \left[ \frac{\curlyE_+}{1 - i (\Delta - \delta \omega) T_2} \int \frac{dz}{L}\ w_0(z) |\Upsilon_+(z)|^2  \right. \nonumber \\
& + |\curlyEtilde_0|^2 \curlyE_+ \int \frac{dz}{L}\ w_0(z) \Lambda^+_+(z) |\Upsilon_0(z)|^2 |\Upsilon_+(z)|^2  \nonumber \\
& \left. + \curlyEtilde_0^2 \curlyE_-^* e^{i \bar{\omega} t} \int \frac{dz}{L}\ w_0(z) \Lambda^-_+(z) \Upsilon_0(z)^2 \Upsilon_-^*(z) \Upsilon_+^*(z) \right] \label{app:eq:dE+dt_general}
\end{align}
\begin{align}
\dot{\curlyE}_- = -\frac{\bar{\ell}}{2} \curlyE_- + \frac{\bar{\alpha}}{2} & \left[ \frac{\curlyE_+}{1 - i (\Delta + \delta \omega) T_2} \int \frac{dz}{L}\ w_0(z) |\Upsilon_-(z)|^2  \right. \nonumber \\
& + |\curlyEtilde_0|^2 \curlyE_- \int \frac{dz}{L}\ w_0(z) \Lambda^-_-(z) |\Upsilon_0(z)|^2 |\Upsilon_-(z)|^2  \nonumber \\
& \left. + \curlyEtilde_0^2 \curlyE_+^* e^{i \bar{\omega} t} \int \frac{dz}{L}\ w_0(z) \Lambda^+_-(z) \Upsilon_0(z)^2 \Upsilon_+^*(z)\Upsilon_-^*(z) \right]. \label{app:eq:dE-dt_general}
\end{align}
\end{widetext}
These three equations will be used to understand the instability threshold. In general, one must first apply the steady-state condition $\dot{\curlyE}_0=0$ to Eq. \ref{app:eq:dE0dt_general} which, together with the Bloch equation relating the field to the inversion, will yield the amplitude of the primary mode $\curlyE_0$ along with the resulting population inversion $w_0(z)$, both as a function of the pumping $w_{eq}$. (The result will be the same as what we calculated for the single-mode solution in Appendix \ref{app:sec:singlemodetheory}.) This information is then used in Eqs.\ \ref{app:eq:dE+dt_general}-\ref{app:eq:dE-dt_general} to determine the minimum level of pumping $w_{eq}$ at which a pair of sidebands with detuning $\dw$ experiences more gain than loss. This is the instability threshold.

So far, we have kept Eqs.\ \ref{app:eq:dE0dt_general}-\ref{app:eq:dE-dt_general} as general as possible to account for arbitrary spatial profiles, GVD, and detuning $\Delta$ between the lasing mode and the peak of the gain spectrum. From here on we will simplify the problem by taking $\bar{\omega}=0$ (zero GVD) and  $\Delta=0$, and apply these conditions to the simplest possible traveling-wave and standing-wave cavities.

\subsection{Traveling-wave cavity}

\begin{figure*}
\includegraphics[scale=0.7]{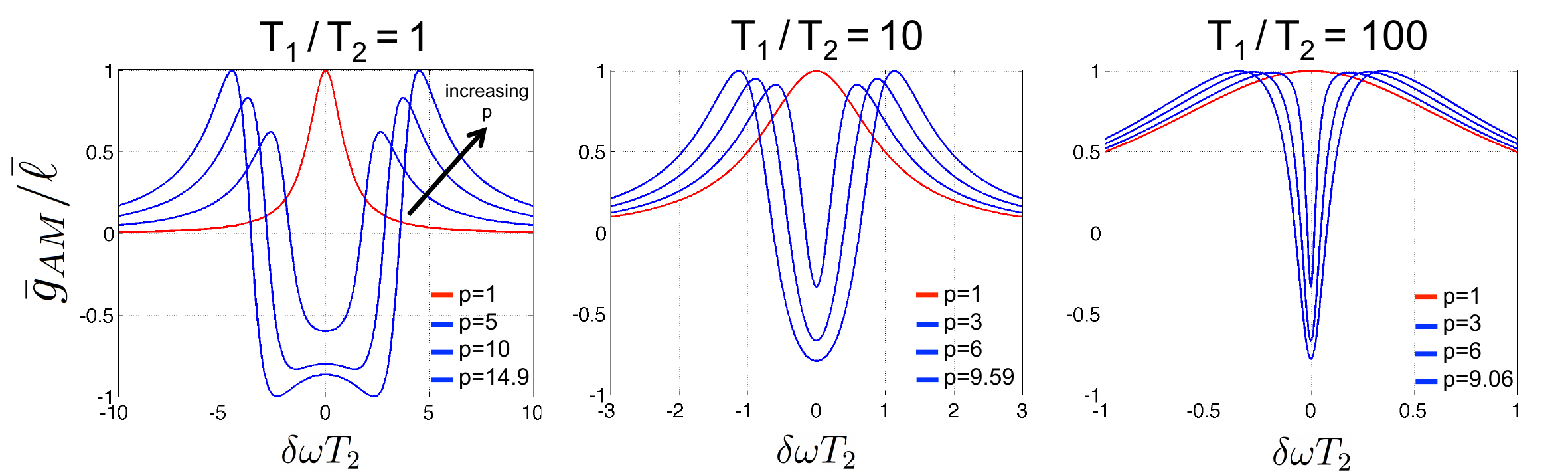}
\caption{\label{app:fig:travelingwave_instabilitycondition} The sideband gain $\bar{g}_{AM}/\bar{\ell}$ of a traveling-wave laser, given in Eq.\ \ref{app:eq:sidebandgain_travelingwave}, is plotted at various pump strengths, for three different values of $T_1/T_2$: 1, 10, and 100. The largest value of $p$ in each plot is equal to the instability threshold given in Eq.\ \ref{app:eq:pRNGH}.}
\end{figure*}

For the traveling-wave laser, the spatial modes are
\begin{equation}
\Upsilon_m(z) = e^{- i k_m z}
\end{equation}
so every point in the cavity sees the same intensity. At and above threshold, the population inversion is everywhere saturated to the threshold inversion, so $w_0$ is independent of $z$. For $\Delta=0$, the inversion is
\begin{equation}
w_0 = w_{th} \equiv \frac{\bar{\ell}}{\bar{\alpha}}
\end{equation}
and the intensity of the primary mode is given by
\begin{equation}
|\curlyEtilde_0|^2=p-1
\end{equation}
where we have made use of the normalized primary mode amplitude $\curlyEtilde_0 \equiv \kappa \sqrt{T_1 T_2} \curlyE_0$, and $p$ is the pumping parameter defined as $p \equiv w_{eq}/w_{th}$. Because $|\curlyEtilde_0|^2$ is independent of $z$, all of the $\Lambda$s are independent of z. Furthermore, since both $w_0$ and the $\Lambda$s are independent of $z$, they can be pulled out of the spatial integrals in Eqs.\ \ref{app:eq:dE+dt_general}-\ref{app:eq:dE-dt_general}. These integrals are then equal to one, where we have used the zero GVD condition $\bar{\omega}=0$ in order for the cross-overlap integral (the last integral in each equation) to equal one. The sideband equations become
\begin{align}
\dot{\curlyE}_+ &= -\frac{\bar{\ell}}{2} \curlyE_+ + \frac{\bar{\alpha} w_{th}}{2} \left[ \frac{\curlyE_+}{1 + i \dw T_2}   +  \Lambda |\curlyEtilde_0|^2 \curlyE_+  +  \Lambda \curlyEtilde_0^2 \curlyE_-^* \right] \label{app:eq:dE+dt_travelingwave} \\
\dot{\curlyE}_- &= -\frac{\bar{\ell}}{2} \curlyE_- + \frac{\bar{\alpha} w_{th}}{2} \left[ \frac{\curlyE_-}{1 - i \dw T_2}   +  \Lambda^* |\curlyEtilde_0|^2 \curlyE_-  +  \Lambda^* \curlyEtilde_0^2 \curlyE_+^* \right] \label{app:eq:dE-dt_travelingwave}
\end{align}
which can be written in matrix form
\begin{equation}
\left(
\begin{array}{cc}
\dot{\curlyE}_+ \\
\dot{\curlyE}_-^*
\end{array}
\right)
=
\left(
\begin{array}{cc}
M_+ & R_+ \\
R_-^* & M_-^*
\end{array}
\right)
\left(
\begin{array}{cc}
\curlyE_+ \\
\curlyE_-^*
\end{array}
\right)
\end{equation}
where
\begin{align}
M_+ &= M_-^* = -\frac{\bar{\ell}}{2} + \frac{\bar{\alpha} w_{th}}{2} \left( \frac{1}{1 + i \dw T_2} + \Lambda |\curlyEtilde_0|^2 \right) \\
R_+ &= R_-^* = \frac{\bar{\alpha} w_{th}}{2} \Lambda |\curlyEtilde_0|^2.
\end{align}
(In the last step, we have finally taken the freedom to choose $\curlyEtilde_0$ to be real, which we can do at this point without loss of generality.)

Now, if we assume a solution of the form $\curlyE_\pm \sim e^{\lambda t}$, we find the two solutions for $\lambda$
\begin{equation}
\lambda = \frac{1}{2} [M_+ + M_-^* \pm \sqrt{(M_+ - M_-^*)^2 + 4 R_+ R_-^*}].
\end{equation}
The net gain seen by each sideband is given by  ${\rm Real}(2 \lambda)$ (the factor of two is for intensity gain rather than amplitude gain), which includes the gain minus the loss. Subtracting off the loss, the gain $\bar{g}$ seen by each sideband is
\begin{equation}
\bar{g} = \bar{\alpha} w_{th} \left[ \frac{1}{1 + (\dw T_2)^2} + \left\{
\begin{array}{cc}
 2\ {\rm Real}(\Lambda) |\curlyEtilde_0|^2 & {\rm ;\ AM} \\
0 & {\rm ;\ FM}
\end{array}
\right.
\right]
\end{equation}
where the two solutions correspond to AM and FM sideband configurations.  Finally, we recognize that the gain is pinned at threshold, so $\bar{\alpha} w_{th} = \bar{\ell}$, and we write down the sideband gain normalized to the loss
\begin{equation}
\frac{\bar{g}}{\bar{\ell}} =  \frac{1}{1 + (\dw T_2)^2} + {\rm Real}(\Lambda) |\curlyEtilde_0|^2 \cdot \left\{
\begin{array}{cc}
 2  & {\rm ;\ AM} \\
0 & {\rm ;\ FM}
\end{array}
\right. .
\end{equation}

When the gain $\bar{g}$ exceeds the loss $\bar{\ell}$, the weak sideband amplitudes experience exponential growth, therefore the single-mode solution becomes unstable. Note that the Lorentzian term is always less than 1. This is a direct result of uniform gain clamping in the traveling-wave laser, which clamps the net gain of the mode at the peak of the Lorentzian to zero, and therefore any mode detuned from the peak will see slightly more loss than gain. FM sidebands therefore never become unstable because they only see the Lorentzian gain. On the other hand, AM sidebands induce a PP and with it a coherent gain term, which can provide enough extra gain on top of the Lorentzian background to allow the sidebands to lase,

\begin{equation}
\frac{\bar{g}_{AM}}{\bar{\ell}} = \frac{1}{1+(\dw T_2)^2} + 2\ {\rm Real}(\Lambda) |\curlyEtilde_0|^2 . \label{app:eq:sidebandgain_travelingwave}
\end{equation}

To get a feel for the sideband gain, we have plotted $\bar{g}_{AM}/\bar{\ell}$ in Fig.\ \ref{app:fig:travelingwave_instabilitycondition} at various pump strengths $p$ for $Z=1$, 10, and 100, where $Z \equiv T_1/T_2$. Graphically, we see that at large enough $p$ sidebands will become unstable. Analytically, it is a simple matter to calculate how hard to pump the laser $p$ before the sidebands appear, starting from Eq.\ \ref{app:eq:sidebandgain_travelingwave}. We start by replacing $|\curlyEtilde_0|^2$ with $p-1$, and note that this substitution must also be made in $\Lambda$, which implicitly varies with $|\curlyEtilde_0|^2$. Then, setting $\bar{g}_{AM}/\bar{\ell}$ equal to one, we can solve a simple quadratic formula for $\dw^2$,
\begin{equation}
(\dw T_2)^2 = \frac{-1 + 3 Z (p-1) \pm \sqrt{[1 - 3 Z (p-1)]^2 - 8 Z^2 p (p-1)}}{2 Z^2}. \label{app:eq:RNGH_deltaw}
\end{equation}
Finally, we must apply some physical reasoning: as $p$ is increased past 1, the sideband gain increases. Right at the moment when the instability threshold is reached, $\dw^2$  must take on a single value. Thus, we set the radical in Eq.\ \ref{app:eq:RNGH_deltaw} to zero and solve for $p$. After solving another simple quadratic equation, we find that
\begin{equation}
p = 5 + \frac{3}{Z} \pm 4\sqrt{1 + \frac{3}{2Z} + \frac{1}{2Z^2}}.
\end{equation}
How do we choose between the plus and minus sign? By plugging this expression for $p$ back into Eq.\ \ref{app:eq:RNGH_deltaw}, it is simple to check that only the plus sign yields real-valued solutions for $\dw$. Thus, we have found the instability threshold, which we denote $p_{RNGH}$,
\begin{equation}
p_{RNGH} = 5 + \frac{3}{Z} + 4\sqrt{1 + \frac{3}{2Z} + \frac{1}{2Z^2}}  \label{app:eq:pRNGH}
\end{equation}
because it is the well-known instability threshold found by Risken and Nummedal (see Eq. 3.10 in \cite{Risken1968a}) and Graham and Haken (see Eq. 7.35 in \cite{Graham1968}). Plugging this value of $p$ into Eq.\ \ref{app:eq:RNGH_deltaw} yields the value of $\dw$ of the sidebands when the instability sets in
\begin{equation}
(\dw_{RNGH} T_2)^2 = \frac{4}{Z^2} + \frac{6}{Z} \left( 1 + \sqrt{1 + \frac{3}{2Z} + \frac{1}{2Z^2}} \right).
\end{equation}
One thing to notice is that in the limit $Z \gg 1$ (transverse relaxation must faster than longitudinal relaxation), the instability threshold $p_{RNGH} \rightarrow 9$ from above and $\dw_{RNGH} T_2 \rightarrow \sqrt{12/Z}$.

\subsection{Standing-wave cavity}
As before, we restrict ourselves to the case $\Delta=0$ and $\bar{\omega}=0$. We will see that calculations for the standing-wave cavity are significantly more complicated than for the traveling-wave cavity. The spatial variation of the primary mode causes the inversion $w_0$ and the coupling $\Lambda$ to both depend on $z$, which makes the integrals more difficult to compute. For this reason, we treat the problem to first order in the primary mode intensity $|\curlyEtilde_0|^2$, which allows us to compute the integrals analytically. However, the theory can be extended to higher order at will, or the integrals can always be computed numerically. 

For the standing-wave laser with perfectly reflecting end mirrors, the spatial profile of each mode is given by
\begin{equation}
\Upsilon_m(z) = \sqrt{2} \cos(k_m z).
\end{equation}
The spatial modulation of the intensity is responsible for the spatial modulation of the population inversion $w_0(z)$, given by Eq.\,\ref{app:eq:w0(z)_standingwave}. This population grating has important consequences. For one, it reduces the power of the laser, which is given by Eq.\,\ref{app:eq:LI}. Secondly, the gain is no longer uniformly clamped by the primary lasing mode, which will allow new modes to lase even in the absence of PPs.

The spatial variation of the primary lasing mode also causes $\Lambda$ to vary with position. In keeping with our approximations, we can expand $\Lambda$ to zeroth order in $|\curlyEtilde_0|^2$ because in our equations $\Lambda$ always multiplies $|\curlyEtilde_0|^2$, so the final result is first order in $|\curlyEtilde_0|^2$. We define the zeroth order expansion of $\Lambda$ to be
\begin{equation}
\chi^{(3)} = \frac{ - (1+i \dw T_2/2) }{(1 + i \dw T_1) (1 + i \dw T_2)^2  } \label{app:eq:Xi3},
\end{equation}
where the symbol $\chi^{(3)}$ was chosen to emphasize that this term now plays the role of a third-order nonlinear coefficient.

We start with the sideband Eqs.\ \ref{app:eq:dE+dt_general}-\ref{app:eq:dE-dt_general}, replace $w_0(z)$ with Eq.\ \ref{app:eq:w0(z)_standingwave}, $\Lambda(z)$ with $\chi^{(3)}$, and keep only terms to first order in $|\curlyEtilde_0|^2$. The resulting equation for the growth of the positive sideband is
\begin{align}
\dot{\curlyE}_+ = -\frac{\bar{\ell}}{2} \curlyE_+ + \frac{\bar{\alpha} w_{th}}{2} & \left[ \frac{1 + \frac{\gamma_D}{2}|\curlyEtilde_0|^2}{1 + i \delta \omega T_2} \curlyE_+ \right. \nonumber \\
& + \chi^{(3)} |\curlyEtilde_0|^2 \curlyE_+ \int \frac{dz}{L}\ |\Upsilon_0(z)|^2 |\Upsilon_+(z)|^2  \nonumber \\
& \left. + \chi^{(3)} \curlyEtilde_0^2 \curlyE_-^* \int \frac{dz}{L}\  \Upsilon_0(z)^2 \Upsilon_-^*(z) \Upsilon_+^*(z) \right],
\end{align}
and a similar equation can be written down for $\dot{\curlyE}_-$. We define the longitudinal overlap integrals
\begin{align}
\Gamma_{\rm self} &=  \int_0^L \frac{dz}{L} |\Upsilon_0(z)|^2 |\Upsilon_+(z)|^2 = 1 \\
\Gamma_{\rm cross} &= \int_0^L \frac{dz}{L}\Upsilon_0(z)^2 \Upsilon_-^*(z) \Upsilon_+^*(z) = 1/2.
\end{align}
The implication is that the self-mixing interaction of a sideband with itself, mediated by the primary mode intensity, is twice as large as the cross-mixing interaction of one sideband generating gain for the other sideband, again mediated by the primary mode intensity. This is true only for the cosine-shaped modes that we have assumed, and the overlap integrals will change when the longitudinal spatial profile changes,  as when the non-unity reflectivity of the facets is taken into account. The sideband equations become
\begin{widetext}
\begin{align}
\dot{\curlyE}_+ &= -\frac{\bar{\ell}}{2} \curlyE_+ + \frac{\bar{\alpha} w_{th}}{2}  \left[ \frac{1 + \frac{\gamma_D}{2} |\curlyEtilde_0|^2}{1 + i \delta \omega T_2} \curlyE_+ + \Gamma_{\rm self} \chi^{(3)} |\curlyEtilde_0|^2 \curlyE_+  + \Gamma_{\rm cross} \chi^{(3)} \curlyEtilde_0^2 \curlyE_-^*  \right] \\
\dot{\curlyE}_- &= -\frac{\bar{\ell}}{2} \curlyE_- + \frac{\bar{\alpha} w_{th}}{2}  \left[ \frac{1 + \frac{\gamma_D}{2} |\curlyEtilde_0|^2}{1 - i \delta \omega T_2} \curlyE_- + \Gamma_{\rm self} {\chi^{(3)}}^* |\curlyEtilde_0|^2 \curlyE_-  + \Gamma_{\rm cross} {\chi^{(3)}}^* \curlyEtilde_0^2 \curlyE_+^*  \right],
\end{align}
\end{widetext}
which we express as
\begin{equation}
\left(
\begin{array}{cc}
\dot{\curlyE}_+ \\
\dot{\curlyE}_-^*
\end{array}
\right)
=
\left(
\begin{array}{cc}
M_+ & R_+ \\
R_-^* & M_-^*
\end{array}
\right)
\left(
\begin{array}{cc}
\curlyE_+ \\
\curlyE_-^*
\end{array}
\right)
\end{equation}
where
\begin{align}
M_+ &= M_-^* = -\frac{\bar{\ell}}{2} + \frac{\bar{\alpha} w_{th}}{2} \left( \frac{1 + \frac{\gamma_D}{2} |\curlyEtilde_0|^2}{1 + i \dw T_2} + \Gamma_{\rm self} \chi^{(3)} |\curlyEtilde_0|^2 \right) \\
R_+ &= R_-^* = \frac{\bar{\alpha} w_{th}}{2} (\Gamma_{\rm cross} \chi^{(3)} |\curlyEtilde_0|^2) .
\end{align}
As we did for the traveling-wave laser, the sideband gain is easily calculated from these two coupled first-order differential equations. Normalizing the gain to the total loss, we find
\begin{align}
\frac{\bar{g}} {\bar{\ell}} &=  \frac{1 + \frac{\gamma_D}{2} |\curlyEtilde_0|^2}{1 + (\dw T_2)^2} \nonumber \\
&+ {\rm Real}[\chi^{(3)}] |\curlyEtilde_0|^2 \cdot \left\{
\begin{array}{cc}
 \Gamma_{\rm self} + \Gamma_{\rm cross} =  \frac{3}{2} & {\rm ;\ AM} \\
 \Gamma_{\rm self} - \Gamma_{\rm cross} =  \frac{1}{2} & {\rm ;\ FM}
\end{array}
\right. , \label{app:eq:standingwave_sidebandgain}
\end{align}
which is equivalent to Eq.\,\ref{eq:standingwave_sidebandgain} after expressing $|\curlyEtilde_0|^2$ in terms of $p$ from Eq.\,\ref{app:eq:LI}. There are two things to notice here. As the laser pumping is increased, the term $\gamma_D |\curlyEtilde_0|^2/2$ grows, and consequently the gain is not clamped at the threshold value. This is due to spatial hole burning, or more precisely, the imperfect overlap of the standing-wave modes together with a finite amount of carrier diffusion.  We view this background gain as a Lorentzian-shape whose amplitude increases with the pumping, and is therefore fully capable of pulling the sidebands above threshold, without any additional PP contribution to the gain.

Secondly, the PP contribution to the gain never vanishes. Even when the sidebands are phased such that an FM waveform is emitted from the laser, there is still a PP within the laser cavity. The reason for this is the imperfect overlap of the two sidebands' spatial modes, which means that at any given position within the cavity, the plus and minus sideband are likely to have different amplitudes. Therefore, even if the two sidebands are phased such that their contributions to the beat note at $\dw$ destructively interfere with each other, the destruction is not perfect. The amplitude of the PP varies with position in the cavity, and in locations where the two sideband amplitudes are equal the PP will not exist, but the spatially averaged effect of the FM PP yields the factor of $1/2$ in Eq.\ \ref{app:eq:standingwave_sidebandgain}. By the same token, sidebands phased for AM will not fully constructively interfere, yielding a factor of $3/2$ for the PP contribution to the gain rather than the factor $2$, as it would be for the traveling-wave laser. 

Depending on the particular values of $T_1$, $T_2$, and $\gamma_D$, either FM or AM sidebands will have a lower instability threshold. This can be determined by solving Eq.\,\ref{app:eq:standingwave_sidebandgain} for the pumping level at which the $\bar{g}=\bar{\ell}$, and gives rise to the three different kinds of instability discussed in Sec.\,\ref{sec:instabilitythreshold} of the main text.

\section{Comparison of theory and data}
\label{app:sec:numerical}

\begin{figure}
\includegraphics[scale=0.97]{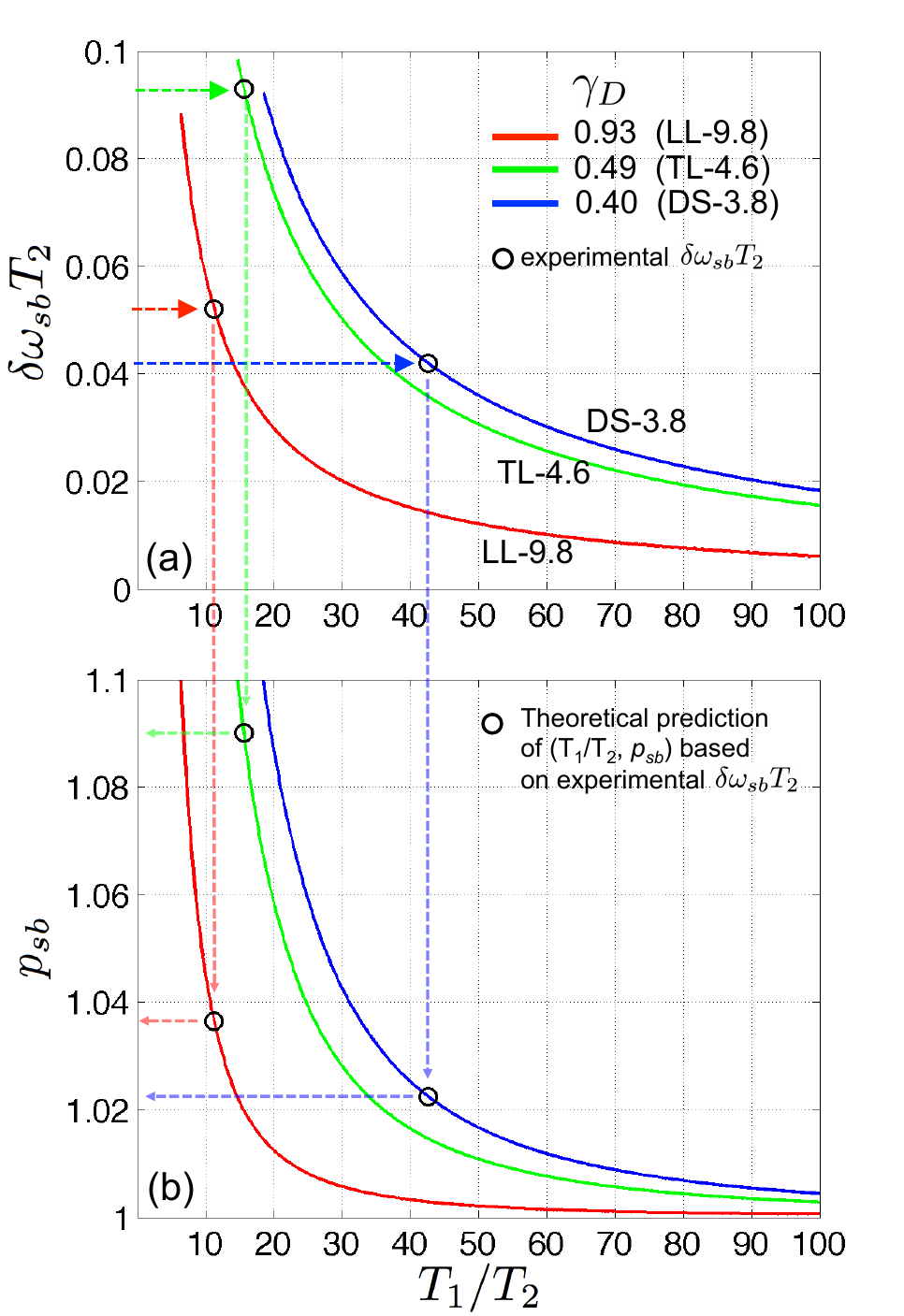}
\caption{\label{app:fig:instabilitythreshold} Numerical solutions of the instability threshold obtained by setting the gain $\bar{g}$ in Eq.\ \ref{app:eq:standingwave_sidebandgain} equal to the loss $\bar{\ell}$, yielding both (a) the sideband separation $\dw_{sb}T_2$ and (b) the pumping $p_{sb}$. The experimentally measured values of $\dwsb T_2$ are compared to the theory to infer $T_1/T_2$, which also gives the theoretical prediction for the instability threshold $p_{sb}$.}
\end{figure}

In this appendix, we take the theory at face value and calculate the predictions of the theory for the three uncoated lasers. We then compare the results with the measurements. (In the main text, we focused on comparing the experimental $\dwsb$ with the theoretical $\dw_{\rm cr}$, a comparison that is more robust against our uncertainty in $\gamma_D$ and $T_1$ and neglect of GVD and $\Delta$.)

\begin{center}
\begin{table}
\resizebox{\columnwidth}{!}{
\begin{tabular}{|l|c|c|c|c|>{\columncolor[gray]{0.8}}c|>{\columncolor[gray]{0.8}}c|c|}
\hline
Device & $T_{\rm up} $ [ps] & $\gamma_D$ & $T_2$ [fs] &  $\dwsb T_2$ & $T_1/T_2$ & $p_{\rm sb}$ &$\Jsb/\Jth$\\ \hline
 LL-9.8 & 0.54 & 0.93 & 81 &  0.052  & 11 & 1.04 & 1.14 \\ \hline
TL-4.6 &  1.7 & 0.49 & 74 &  0.093  & 16 & 1.09 & 1.17 \\ \hline
DS-3.8 &  1.74 & 0.40 & 43 &  0.042 & 43 & 1.02 & 1.12 \\
\hline
\end{tabular}}
\caption{Summary of the input parameters and the output theoretical predictions for the three uncoated devices. $T_{\rm up}$ is used to calculate $\gamma_D$. $T_2$ and $\dwsb T_2$ are measured quantities, from which the columns $T_1/T_2$ and $p_{\rm sb}$ are calculated from the theory, as explained in Fig.\,\ref{app:fig:instabilitythreshold}. The quantity $p_{\rm sb}$ clearly underestimates the measured $\Jsb/\Jth$ for reasons discussed in the text. \label{tab:expvstheory} }
\end{table}
\end{center}

Because the theory assumes end mirrors with unity reflectivity, we can only expect Eq.\ \ref{app:eq:standingwave_sidebandgain} to apply reasonably well to the uncoated QCLs. For each device, $\gamma_D$ is calculated using the theoretical value of $T_{\rm up}$ (calculated from the bandstructure) and the diffusion constant $D = 77$ cm$^2$/s \cite{Faist2013}, giving $\gamma_D = 0.4$ (DS-3.8), 0.49 (TL-4.6), and 0.93 (LL-9.8). For these large values of $\gamma_D$, the incoherent gain increases rapidly with the pumping, and we find from Eq.\ \ref{app:eq:standingwave_sidebandgain} that the FM instability will have a lower threshold than the AM instability, regardless of the value of $T_1$. The gain recovery time $T_1$ of each QCL is not as easily calculable as $T_{\rm up}$ because it depends on a few other time constants of the active region, such as the escape time of the electron from one injector region to the next active region. Therefore, we treat $T_1$ as a variable and calculate the instability threshold $p_{\rm sb}$ and sideband spacing $\dw_{\rm sb}$ as a function of $T_1$. The resulting curves are shown in Fig.\ \ref{app:fig:instabilitythreshold}, and a summary of all input and output parameters is given in Table \ref{tab:expvstheory}. By comparing the curves with the measured values of $\dwsb$, we can deduce the values $T_1=1.83$ ps (DS-3.8),  1.15 ps (TL-4.6), and 0.91 ps (LL-9.8). For these values of $T_1$, the theory predicts an instability threshold of $p_{\rm sb}=1.02$ (DS-3.8), 1.09 (TL-4.6), and 1.04 (LL-9.8). It is encouraging that these fitted values of $T_1$ are close to the accepted value of the QCL gain recovery time, which has been shown by pump-probe experiments \cite{Choi2008,Choi2009} and theory \cite{Talukder2011} to be around 2 ps. However, the predicted $p_{\rm sb}$ is significantly lower than the measured values $J_{\rm sb}/J_{\rm th}=1.12$ (DS-3.8), 1.17 (TL-4.6), and 1.14 (LL-9.8), and the discrepancy is made worse by the fact that $J/J_{\rm th}$ is likely an underestimate of $p$ (see the discussion in Appendix \ref{app:sec:intracavitypower}). The fact that the theory underestimates the instability threshold is perhaps not surprising, as we have only made sure that one of the two necessary conditions for sideband oscillation is satisfied (gain, not phase). Our neglect of the current inhomogeneity also contributes to the underestimation, as discussed in Appendix \ref{app:sec:singlemodetheory}. We hope that future work which accounts for the detuning $\Delta$, the detuning between the lasing mode and the cold cavity mode it occupies, the GVD, and the current inhomogeneity can accurately predict the instability threshold, which would be a milestone in the understanding of lasers, and also yield a novel laser characterization method of lifetimes and diffusion rates by comparing measured values of $p_{\rm sb}$ and $\dw_{\rm sb}$ to an established theory.

\begin{acknowledgements}

This work was supported by the DARPA SCOUT program through grant number W31P4Q-16-1-0002. We acknowledge support from the National Science Foundation under awards ECCS-1230477, ECCS-1614631, and ECCS-1614531. This work was performed in part at the Center for Nanoscale Systems (CNS), a member of the National Nanotechnology Coordinated Infrastructure (NNCI), which is supported by the National Science Foundation under NSF award no. 1541959. CNS is part of Harvard University. We gratefully acknowledge the Office of Naval Research (ONR) for assistance in developing the 3.85 $\mu$m QCL used in this study. The Lincoln Laboratory contribution is based upon work supported by the Assistant Secretary of Defense for Research and Engineering under Air Force Contract No. FA8721-05-C-0002 and/or FA8702-15-D-0001. Any opinions, findings, conclusions or recommendations expressed in this material are those of the authors and do not necessarily reflect the views of the Assistant Secretary of Defense for Research and Engineering. Benedikt Schwarz was supported by the Austrian Science Fund (FWF) within the doctoral school Solids4Fun (W1243) and the project NanoPlas (P28914-N27). Tobias Mansuripur thanks Dmitry Kazakov and Marco Piccardo for measurements of the beat note in the dense state, Carlos Stroud for his perspective on laser instability research in the 1980s, and Jacob Khurgin for the remark that initiated this research.


\end{acknowledgements}

\bibliography{library.bib}

\end{document}